\documentclass[12pt]{article}

\usepackage{amsmath}
\usepackage{amsfonts}
\usepackage{amssymb}
\usepackage[nosort]{cite}
\usepackage{graphics,graphicx,tikz}
\usepackage{soul}
\usetikzlibrary{patterns}

\numberwithin{equation}{section} 
\usepackage[linktocpage]{hyperref}
\hypersetup{
	colorlinks=true,
	linkcolor=blue,
	filecolor=magenta,      
	urlcolor=blue,
	citecolor=blue
}
\def\beq{\begin{equation}}
\def\eeq{\end{equation}}
\def\bea{\begin{align}}
\def\eea{\end{align}}
\newcommand\txt[1]{{#1}}

\usepackage{color}

\begin{document}
\begin{titlepage}
\hfill \hbox{CERN-TH-2022-169}
\vskip 0.1cm
\hfill \hbox{NORDITA 2022-071}
\vskip 0.1cm
\hfill \hbox{QMUL-PH-22-32}
\vskip 0.1cm
\hfill \hbox{UUITP-42/22}
\vskip 0.1cm
\begin{flushright}
\end{flushright}
\vskip 1.0cm
\begin{center}
{\Large \bf Classical Gravitational Observables from the  Eikonal Operator}
\vskip 1.0cm {\large  Paolo Di Vecchia$^{a, b}$, Carlo Heissenberg$^{c,b}$,
Rodolfo Russo$^{d}$, \\
Gabriele Veneziano$^{e, f}$ } \\[0.7cm]

{\it $^a$ The Niels Bohr Institute, Blegdamsvej 17, DK-2100 Copenhagen, Denmark}\\
{\it $^b$ NORDITA, KTH Royal Institute of Technology and Stockholm University, \\
Hannes Alfv{\'{e}}ns v{\"{a}}g 12, SE-11419 Stockholm, Sweden  }\\
 {\it $^c$ Department of Physics and Astronomy, Uppsala University,\\ Box 516, SE-75120 Uppsala, Sweden}\\
{\it $^d$ Queen Mary University of London, Mile End Road,\\ E1 4NS London, United Kingdom}\\
{\it $^e$ Theory Department, CERN, CH-1211 Geneva 23, Switzerland}\\
{\it $^f$Coll\`ege de France, 11 place M. Berthelot, 75005 Paris, France}
\end{center}
\begin{abstract}
We propose two possible eikonal operators encoding the effects of 
classical radiation as coherent states of gravitons and show how to 
compute from them different classical observables. In the first proposal, 
only genuinely propagating gravitons are included, while, in the second, 
zero-frequency modes are added in order to recover the effects of a static 
gravitational field. We first calculate the radiated energy-momentum and 
the change in each particle's momentum, or impulse, to 3PM order finding 
agreement with the literature. We then calculate the angular momentum of 
the gravitational field after the collision. In order to do so, we adapt 
the method of reverse unitarity to the presence of derivatives in the 
operators describing the angular momentum and reproduce the result 
of~\cite{Manohar:2022dea} obtained by resumming the small-velocity 
expansion. As a new application, we derive also the variation in each 
particle's angular momentum up to 3PM: calculating separately field and 
particle contributions allows us to check the balance laws explicitly. We 
also show how the eikonal operator encodes the linear-response formula of 
Bini-Damour by deriving the linear radiation-reaction contribution to the 
transverse impulse at 4PM.
\end{abstract}
\end{titlepage}

\section{Introduction}
Classical gravitational interactions are mediated by the exchange of a large number of very soft quanta, and the eikonal exponentiation makes this intuitive notion precise, resumming the contributions due to infinitely many graviton exchanges into a rapidly oscillating phase \cite{Amati:1987wq,Amati:1990xe,Kabat:1992tb,Akhoury:2013yua,Bjerrum-Bohr:2018xdl}. On the other hand, the final state of a collision event also includes gravitational waves, which can be described by a  superposition of graviton states \cite{Ciafaloni:2015xsr,Ciafaloni:2018uwe,Cristofoli:2021jas}. In this letter, we establish a link between these two phenomena by means of an eikonal operator, describing how the colliding objects evolve due to the mutual interaction, emit classical radiation and how this backreacts on their trajectories \cite{Amati:1990xe,Ciafaloni:2015xsr,Ciafaloni:2018uwe,Cristofoli:2021jas,Addazi:2019mjh,Damgaard:2021ipf,DiVecchia:2022owy,DiVecchia:2022nna}. In the Post-Minkowskian (PM) approach, this is achieved at leading order by combining the $2\to2$ and $2\to3$ amplitudes into an exponential operator that dictates the collision's final state. 

This is sufficient to calculate all classical observables up to 3PM i.e.~$\mathcal{O}(G^3)$ order~\cite{Bern:2019nnu,Bern:2019crd,Kalin:2020fhe,Parra-Martinez:2020dzs,DiVecchia:2020ymx,Herrmann:2021lqe,DiVecchia:2021bdo,Herrmann:2021tct,Bjerrum-Bohr:2021vuf,Bjerrum-Bohr:2021din,Brandhuber:2021eyq,Manohar:2022dea,Jakobsen:2022psy,Kalin:2022hph}. We derive to 3PM precision the changes in each particle's momentum (the impulse) \cite{Herrmann:2021tct}, which include both the transverse and the longitudinal components with respect to the initial directions of motion, the linear \cite{Herrmann:2021lqe} and angular momentum \cite{Manohar:2022dea} of the gravitational radiation field and, as a novel application, the changes in each particle's angular momentum.
In this way, we can explicitly check the corresponding balance laws to the same order.

More in detail, we will actually present two, in principle distinct, eikonal operators: one obtained by an operator dressing of the (connected) $T$-matrix, based on the standard version of Weinberg's soft-graviton theorem \cite{Weinberg:1964ew,Weinberg:1965nx,Weinberg:1972kfs,DiVecchia:2022nna}, and one obtained by dressing the whole $S$-matrix, $S=1+iT$, including the forward part \cite{DiVecchia:2022owy}. We show that the linear momenta of the gravitational field and the particles are the same in both formulations although certain terms emerge from different ingredients of the respective operators as in the case of the impulse, and our results agree with \cite{Herrmann:2021lqe,Herrmann:2021tct}. The difference between the two approaches emerges in the results for the angular momenta.

At the technical level, the main contribution to the radiative quantities we study can be written in terms of ``cut'' two-loop integrals that we calculate by using the technique of reverse unitarity \cite{Anastasiou:2002yz,Anastasiou:2002qz,Anastasiou:2003yy,Herrmann:2021lqe,Herrmann:2021tct} and the method of differential equations \cite{Parra-Martinez:2020dzs,DiVecchia:2021bdo,Herrmann:2021tct}. When applying this approach to the angular momentum, whose expression contains derivatives with respect to the momenta, one encounters a new feature: the possibility for the derivatives to act on the delta functions enforcing the on-shell conditions. We show that the approach mentioned above can be implemented in a non-ambiguous way and that the final results always depend only on on-shell data. 

In addition to conservative and radiative effects, the second formulation of the eikonal operator allows us to access also ``static'' contributions, e.g.~to the angular momentum losses, which are sensitive to the prescription adopted to approach the $\omega=0$ singularity of the graviton spectrum \cite{DiVecchia:2022owy}. These static contributions depend on the Bondi supertranslation frame one adopts \cite{Veneziano:2022zwh}, and employing the Feynman $-i0$ prescription \cite{DiVecchia:2022owy} corresponds to adopting the Bondi frame where the linear-response formula of Bini and Damour \cite{Bini:2012ji,Bini:2021gat} holds. In this frame, our result for the total angular momentum loss, complete of radiative and static terms, agrees with \cite{Damour:2020tta} to 2PM order and with \cite{Manohar:2022dea} to 3PM order.

We also show that
the eikonal operator can naturally account for the linear radiation reaction (RR) contribution to the impulses as derived in a classical GR approach in~\cite{Bini:2012ji,Bini:2021gat}. This linear-response formula was used recently in~\cite{Manohar:2022dea} to obtain the transverse component of the impulse at ${\cal O}(G^4)$ and we show that the same result follows from the eikonal operator that includes the static contributions.

The paper is organized as follows. In Section~\ref{sec:eikop12} we propose our formulation of the eikonal operator, including either only radiative modes (Subsection~\ref{sec:form1}) or both radiative and static modes (Subsection~\ref{sec:form2}). Section~\ref{sec:observables} is then devoted to the explicit calculations of classical observables from the eikonal operator up to 3PM order: the total radiated momentum, the impulse for each particle, the total angular momentum of the gravitational field and the change in each particle's angular momentum. 
In Section~\ref{sec:outlook} we illustrate how the 4PM radiation-reaction part \cite{Manohar:2022dea} of the transverse impulse also follows from the formalism, and comment on possible future applications aimed to obtain other contributions at the same order.

\section{Two Proposals for the Eikonal Operator}
\label{sec:eikop12}

We focus on the gravitational scattering of two scalar particles with masses $m_1$ and $m_2$ in the PM approximation. 
All momenta are conventionally regarded as outgoing, so that $-p_1$ and $-p_2$ are the classical momenta of the incoming particles, with $p_1^2+m_1^2=0$, $p_2^2+m_2^2=0$. The initial configuration of the collision is described by a state with no radiation. Following~\cite{Kosower:2018adc}, and introducing the shorthand notation
	$\int_{p_{i}} = \int \frac{d^Dp_{i}}{(2\pi)^D}\,2\pi\theta(p_i^0)\delta(p_i^2+m_i^2)$
for $i=1,2$ (and similarly for $p_{3,4}$),
we thus indicate the in-state as
\begin{equation}
	| \psi \rangle = \int_{-p_1}\int_{-p_2}
	\Phi_1 (-p_1) \Phi_2 (-p_2)  e^{i b_1\cdot p_1 + i b_2\cdot p_2} |- p_1 ,- p_2,0 \rangle\,.
	\label{DO1}
\end{equation}
The wavepackets $\Phi_i(-p_i)$ are peaked around the classical incoming momenta (which, with a slight abuse of notation, we indicate as $-p_i$), while $b=b_1-b_2$ is the impact parameter and is orthogonal to $p_{1,2}$.
The ket label ``$0$''  emphasizes that there are no gravitons in the initial state.
The basic principle of the eikonal approach is that it is possible to describe the final state, after a gravitational scattering at large $b$, by an exponential that includes all classical effects. 

In the elastic case, the final state contains just two scalar particles, like the initial one, and can be written as~\cite{Cristofoli:2021jas} (see in particular Eq.~(4.7) of that reference)
\begin{equation}\label{eq:eikel}
\begin{split}
S_\text{el} |\psi\rangle & = \int_{p_3} \int_{p_4}
e^{-ib_1\cdot p_4-ib_2\cdot p_3}\,  \int\frac{d^D \tilde{Q}}{(2\pi)^D} \Phi_1(p_4-\tilde{Q})\,\Phi_2(p_3+\tilde{Q}) \\ & \times \int d^D\!\tilde{x}\, e^{i(b_1-b_2-\tilde{x})\cdot\tilde{Q}} (1+2i\Delta(\sigma,b_e))\, e^{2 i \delta(\sigma,b_e)} |p_3,p_4,0\rangle\,.
\end{split}
\end{equation}      
Here $\sigma=-\frac{p_1\cdot p_2}{m_1 m_2}=-\frac{p_3\cdot p_4}{m_1 m_2}$ and $2\delta$ is the classical eikonal while $\Delta$ is a quantum remainder, see \cite{DiVecchia:2021bdo} and references therein.
In terms of the Fourier transform of the $2\to2$ amplitude $\mathcal M$ to impact parameter space, $\operatorname{FT}\left[\mathcal M\right]\equiv \widetilde{\mathcal{M}}$,
\begin{equation}\label{FT2}
	\widetilde{\mathcal M}(\sigma,b_e) = \int \frac{d^Dq}{(2\pi)^D}\,2\pi\delta(2p_1\cdot q - q^2)2\pi\delta(2p_2\cdot q + q^2)\,e^{ib_e\cdot q}\mathcal M(\sigma,q^2)\,,
\end{equation}
it satisfies
\begin{equation}\label{eikEL}
	1+2i\widetilde{\mathcal{M}}(\sigma,b_e)
	= \left(1+2i\Delta(\sigma,b_e)\right)e^{2i\delta(\sigma,b_e)}\,.
\end{equation}
We will neglect $\Delta$ since it will not play an important role in this work. In Eq.~\eqref{eq:eikel}, the impact parameter $b_e$ is defined as the projection of ${\tilde{x}}$ orthogonal to $p_4-\tilde{Q}/2$ and $p_3+\tilde{Q}/2$, so it implicitly depends on $\tilde{Q}$ and, as discussed in detail in~\cite{Cristofoli:2021jas}, this yields the expected relation between the impact parameter $b$ defining the initial angular momentum and the eikonal impact parameter $b_e$, see \eqref{tildebeb}.
Up to $\mathcal{O}(G^3)$, one has $2\delta \simeq 2\delta_0+ 2\delta_1 + 2\delta_2$. The 1PM ($2\delta_0$) and 2PM ($2\delta_1$) orders are real, so that $e^{2i\delta_0+2i\delta_1}$ is a phase~\cite{KoemansCollado:2019ggb,Cristofoli:2020uzm}. Explicitly:
\begin{equation} \label{delta1-2PM}
	2\delta_0
	=
	\frac{2G m_1m_2 \Gamma(-\epsilon)}{(\pi b^2_e)^{-\epsilon}}\,\frac{\sigma^2-\frac{1}{2(1-\epsilon)}}{\sqrt{\sigma^2-1}}\,,
	\quad
		2\delta_1
		=
	\frac{3\pi G^2 m_1m_2(m_1+m_2)}{4b_e}\,\frac{5\sigma^2-1}{\sqrt{\sigma^2-1}}\,,
\end{equation}        
where we kept $D=4-2\epsilon$ in $2\delta_0$ to regulate the Coulomb IR divergence. Instead, the 3PM order $2\delta_2$ develops a (positive) IR divergent imaginary part showing that the elastic process is exponentially suppressed as inelastic channels open up. 
In order to recover unitarity we thus need to consider radiation in the final state.

Building on several approaches~\cite{Ciafaloni:2018uwe,Addazi:2019mjh,Damgaard:2021ipf,Cristofoli:2021jas,DiVecchia:2022owy,DiVecchia:2022nna}, we write the final state as
\begin{equation}\label{eikopg}
\begin{split}
	S|\psi\rangle
&\simeq
\int_{p_3}\int_{p_4}
e^{-ib_1\cdot p_4-ib_2\cdot p_3}\,  \int\frac{d^DQ_1}{(2\pi)^D} \int\frac{d^DQ_2}{(2\pi)^D} 
\, \Phi_1(p_4-Q_1)\,\Phi_2(p_3-Q_2) \\
&\times \int d^D\!x_1 \int d^D\!x_2
\,e^{i (b_1 - x_1)\cdot Q_1+ i(b_2-x_2)\cdot Q_2}\, e^{2i\hat{\delta}(x_1,x_2)} |p_3,p_4,0\rangle\,.
\end{split}
\end{equation}
The momenta $Q_i$ thus enter the relation between initial and final states, $p_1+p_4=Q_1$ and $p_2+p_3=Q_2$. 
The operator $2\hat{\delta}$ lives in the Fock space of the physical gravitons and, as we shall see,
it contains two main ingredients. The first is the (real) c-number  eikonal phase derived from the elastic $2\to2$ amplitude \eqref{eikEL}. The second is an inelastic part describing radiation that involves amplitudes with graviton emissions 
and the creation/annihilation operators $a_i(k)$, $a_i^\dagger(k)$, with $i$ the polarization index, obeying canonical commutation relations
	$2\pi\theta(k^0)\delta(k^2)[a_i(k),a_j^\dagger(k')]
	=
	(2\pi)^D \delta^{(D)}(k-k')\delta_{ij}$.
As anticipated, we shall detail two somewhat different proposals for $2\hat\delta$. Both involve the Fourier transform of the connected  on-shell $2\to3$ amplitude $\mathcal A^{\mu\nu}$ in the classical limit (see~\cite{Goldberger:2016iau,Luna:2017dtq,Mogull:2020sak} for the tree-level result),\footnote{In analogy with our treatment of the elastic amplitude, $\mathcal A^{\mu\nu}$ contains, in principle, conservative loop corrections but no absorptive ones due to inelastic on shell intermediate states.}  
\begin{equation}\label{A}
	\tilde{\mathcal{A}}^{\mu\nu}(x_1,x_2,k) = \int \frac{d^Dq_1}{(2\pi)^{D-2}}\,\delta(2 p_1\cdot q_1-q_1^2) \delta(2 p_2\cdot q_2-q_2^2) e^{ix_1\cdot q_1+ix_2\cdot q_2}\mathcal A^{\mu\nu}(q_1,q_2,k)\,,
\end{equation}
where $q_1+q_2+k=0$. The following transformation property under translations:
\begin{equation}\label{translation}
	x^\mu_{1,2}\to x^\mu_{1,2}+a^\mu,\qquad
	\tilde{\mathcal{A}}^{\mu\nu} \to  e^{-ia\cdot k}\tilde{\mathcal{A}}^{\mu\nu}\,,
\end{equation} 
will be useful in the calculation of the angular momenta in Sect. \ref{sec:Radiative-modes}.

\subsection{Eikonal operator without static modes}
\label{sec:form1}

A first approach is to define the eikonal operator in~\eqref{eikopg} by including a coherent superposition of graviton states~\cite{Cristofoli:2021jas}, 
\begin{equation}
\label{eikope}
e^{2i\hat{\delta}(x_1,x_2)}=  \int\!\frac{d^D \tilde{Q}}{(2\pi)^D} \int\! d^D\tilde{x} \, e^{-i\tilde{Q}(\tilde{x}-x_1+x_2)}
e^{2i\delta_s(b_e)} e^{i\int_{k}\left[\tilde{\mathcal{A}}_j(x_1,x_2,k) a_j^\dagger(k)+\tilde{\mathcal{A}}_j^\ast(x_1,x_2,k) a_j(k)\right]}\,,
\end{equation}
where the polarization index $j$ is summed over and
	$\int_{k} = \int \frac{d^Dk}{(2\pi)^D}\,2\pi\theta(k^0)\delta(k^2)\,.$
Note that neglecting radiation, i.e.~setting $\tilde{\mathcal{A}}$ to zero in the eikonal operator \eqref{eikope}, Eq.~\eqref{eikopg} reduces to the elastic final state \eqref{eq:eikel}.

Eq.~\eqref{eikope} includes the region of small $k$, say $k^0<\omega^\ast$, where Weinberg's soft theorem~\cite{Weinberg:1965nx} for ${\mathcal{A}}^{\mu\nu}$ applies. In this region, the last two factors in \eqref{eikope} reduce to the small-frequency eikonal operator discussed in \cite{DiVecchia:2022nna}: letting $\int_k^{\omega^\ast}\equiv  \int_k \theta(\omega^\ast-k^0)$,
\begin{equation} \label{KF}
	S_{s.r.} = e^{2i\delta_s}  e^{\int_k^{\omega^\ast}\left[
		w_j a_j^\dagger 
		- w_j^{\ast} a_j
		\right]} 
	= e^{2i\delta_s} e^{\int_k^{\omega^\ast} \left[
		w_j^\text{out} a_j^\dagger 
		- w_j^{\text{out}\ast} a_j
		\right]}
	e^{\int_k^{\omega^\ast} 
		\left[
		w_j^\text{in} a_j^\dagger 
		- w_j^{\text{in}\ast} a_j
		\right]}\,,
\end{equation} 
where 
$	w_j (k) = \varepsilon^{\ast}_{j, \mu\nu}(k) \sum_{n} \sqrt{8\pi G}\, p_n^\mu p_n^\nu /p_n\cdot k$,
and similarly for $w_j^{\text{out/in}}$ but restricting $n$ to final/initial states. Eq.~\eqref{KF} shows that introducing the operator $S_{s.r.}$ is equivalent to performing soft dressings  \cite{Kulish:1970ut,Choi:2017ylo,Mirbabayi:2016axw,Hannesdottir:2019opa} of the initial and final states.

After exponentiation, the initial and the final momenta in the scattering process differ by classical corrections weighted by $G$, rather than by quantum correction weighted by $\hbar$. Thus we need to spell out the dependence of the objects in~\eqref{eikope} on the external momenta. The choice of~\cite{Cristofoli:2021jas} was to identify all momenta with those of the final states, while here we take a ``democratic'' prescription. For instance the phase $2\delta_s(b_e)$ is defined from the eikonal phase $\operatorname{Re}2\delta$ \cite{DiVecchia:2021bdo} (see \eqref{eikEL} above) by
\begin{equation}\label{deltael}
	2\delta_s(b_e)
	=
	\frac12\left[
	\operatorname{Re}2\delta(\sigma_{12},b_e)+\operatorname{Re}2\delta(\sigma_{34},b_e)
	\right].
\end{equation}
The symmetrization between incoming $\sigma_{12}=-\frac{p_1\cdot p_2}{m_1 m_2}$ and outgoing $\sigma_{34}=-\frac{p_3\cdot p_4}{m_1 m_2}$ momenta reproduces the classical Coulomb divergences due to the hard external particles~\cite{Weinberg:1965nx}, and is relevant only to $\mathcal O(G^4)$, i.e.
	$2\delta_s(b_e)
	=
	\operatorname{Re}2\delta(\sigma,b_e)+\mathcal O(G^4)$,
so we will need it in explicit calculations only in Section~\ref{sec:outlook}.

Similarly, we take $\tilde{\mathcal{A}}^{\mu\nu}$ to depend on $\tilde p_1$ and $\tilde p_2$,
\begin{equation}\label{tildep}
  \tilde p_1 = \frac12(p_4-p_1)
  =p_4-\frac{Q_1}{2}\,,\qquad
  \tilde p_2 = \frac12(p_3-p_2)
  =p_3-\frac{Q_2}{2}\,,
\end{equation}
rather than on $p_1$ and $p_2$. As we will see this is again relevant for radiation-reaction effects at $\mathcal{O}(G^4)$.\footnote{Both the phase and the operator part of \eqref{eikopg} will be treated differently in the alternative formulation that reproduces the static contributions to the angular momenta~\cite{Damour:2020tta,Manohar:2022dea,DiVecchia:2022owy}.} Notice that we do not include the imaginary part of $2\delta_2$, since it is automatically generated by the operatorial part after normal ordering \cite{DiVecchia:2022nna}.\footnote{This is another slight difference with the approach~\cite{Cristofoli:2021jas} where the imaginary part is written explicitly. When evaluated at the classical stationary point (see~\eqref{Saddlev1}) the two approaches are equivalent, but in order to calculate classical observables that involve derivatives it is important to start from~\eqref{eikope} and impose~\eqref{Saddlev1} only at the end.} This shows that the damping of the elastic amplitude (which diverges in the $\epsilon \to 0$ limit) is due to the emission of soft-gravitons \cite{DiVecchia:2022nna}. As we will see, this formulation of the final state yields consistent radiative observables up to 3PM.

It is instructive to rewrite Eq.~\eqref{eikopg} after the change of variables $x_1 = \lambda+\frac{x}{2}$, $x_2 = \lambda-\frac{x}{2}$, $Q_1 = Q-\frac{P}{2}$, $Q_2 = -Q-\frac{P}{2}$. Then $P$ represents the total momentum lost by the particles to the gravitational field. In fact, $\tilde{\mathcal{A}}$ only depends on $\lambda$ via an overall $e^{-i\lambda\cdot k}$ and therefore $P$ is equal to the sum of the momenta of the emitted gravitons $P=\sum_m k_m$, as one can see by expanding the last exponential in~\eqref{eikope} to a generic order $(a^{\dagger})^M$ and then carrying out the integral over $\lambda$.

Since $2i\hat{\delta}$ in Eq.~\eqref{eikope} is manifestly Hermitian, the eikonal operator~\eqref{eikopg} would be trivially unitary were it not for the various integrations present in it, in particular the above-mentioned one that enforces energy-momentum conservation. However, we expect unitarity to hold only up to corrections that are higher order either in the $\hbar$-expansion or, less pretentiously, in the PM (loop) expansion.
For instance, one can use this approach to check unitarity at the classical level,\footnote{We checked that all rapidly oscillating phases cancel at the stationary point, but not that the overall coefficient is one up to $\hbar$ corrections. It would be interesting to come back to this point.} i.e. 
$\langle \psi| S^\dagger S |\psi\rangle \simeq \langle \psi| \psi\rangle$.
To this end, we use~\eqref{eikopg} to write explicitly the l.h.s. of the above equation, and estimate the integrals by solving the stationary phase conditions. We find that the integrated variables in $S |\psi\rangle$ must be equal to those in $\langle \psi| S^\dagger$ and that, for $i=1,2$,
\begin{subequations}
\label{Saddlev1}
\begin{align}
\label{saddle1}
(x_i - b_i)_{\mu} &=
\frac{\partial 2\delta_s(b_e)}{\partial Q_i}
-i\int_{k}\tilde{\mathcal{A}}^\ast(x_1,x_2,k)\frac{\overset{\leftrightarrow}{\partial}}{\partial Q_i^\mu}\tilde{\mathcal{A}}(x_1,x_2,k)
  \,, \\ 
\label{saddle2}
Q_{i \,\mu} &=
(-1)^{i+1} \tilde{Q}_{\mu} -i\int_{k}\tilde{\mathcal{A}}^\ast(x_1,x_2,k)\frac{\overset{\leftrightarrow}{\partial}}{\partial x_i^\mu}\tilde{\mathcal{A}}(x_1,x_2,k)  \\
\label{saddle3}
\tilde{x}_\mu &= (x_1-x_2)_\mu  +\frac{\partial 2\delta_s(b_e)}{\partial \tilde{Q}^\mu}\;,~~
\tilde{Q}_\mu = \frac{\partial 2\delta_s(b_e)}{\partial \tilde{x}^\mu}\;,~
\end{align}
\end{subequations}
with $2f\overset{\leftrightarrow}{\partial}g =f{\partial}g-g{\partial}f$.
For convenience, here and in the following we suppress contractions between five-point amplitudes (unless written  otherwise), letting 
\begin{equation}\label{suppressedindices}
	\mathcal A\, \mathcal A' = \mathcal A_{\mu\nu}\, \mathcal A'^{\mu\nu}-\frac{1}{D-2}\,\mathcal A^{\mu}_\mu \mathcal A'^\nu_\nu,
\end{equation}
(and similarly for $F^{\mu\nu}$ below).
The saddle-point equations for the elastic case \eqref{eq:eikel} can be obtained from \eqref{Saddlev1} by formally setting $\mathcal{\tilde{A}}$ to zero.

We witness here the key idea of the eikonal approach, that the classical values of $Q_i$ and $x_i$ can be determined by stationary-phase conditions.
Using \eqref{eikopg} and \eqref{Saddlev1}, one can calculate expectation values of classical observables $O$ by evaluating $\langle \psi| S^\dagger O S |\psi\rangle$ at the stationary point. For the impulse, the operator insertion simply multiplies the integrand by a factor of the momentum (see \eqref{Pformula} below). Considering the angular momenta of both the gravitational field and the scalar particles, instead, it also involves derivatives.  For instance, the mechanical angular momentum of particle 1 before (after) the scattering is obtained by inserting $-i p_{1[\alpha}\frac{\partial }{\partial p_1^{\beta]}} $ ($-i p_{4[\alpha}\frac{\partial }{\partial p_4^{\beta]}}$). Here and below, all formulas for particle 2 can be obtained by interchanging $1\leftrightarrow2$ and $3\leftrightarrow4$ in those for particle 1.

\subsection{Eikonal operator including static modes}
\label{sec:form2}

So far, we worked by exponentiating the $2\to3$ amplitude as in \eqref{eikope}, which includes Weinberg's limit of soft (but non vanishing) graviton momenta. 
We now discuss how to also take into account additional effects associated to static or ``zero-frequency'' modes \cite{DiVecchia:2022owy}, which will be instrumental in including the angular momentum of the non-dynamical gravitational field~\cite{Damour:2020tta,Manohar:2022dea,DiVecchia:2022owy}. This formulation also provides a natural separation between conservative and non-conservative contributions treating all RR effects on the same footing. 

Let us first isolate in~\eqref{eikope} the contribution of the infinitesimally soft gravitons, $k^0 <\omega^\ast$. The scale $\omega^\ast$ is useful only in the intermediate steps to define the integrals of Section~\ref{sec:statcont} and we will take $\omega^\ast\to 0$ at the end. We perform a soft dressing not only of the connected amplitudes ($T$-matrix elements), but of the full $S$-matrix including the disconnected part \cite{DiVecchia:2022owy}. This amounts to incorporating also diagrams where the graviton is emitted by a ``straight line'' and thus carries zero energy, and requires us to introduce the $-i0$ prescription in the of the soft factors:
\begin{equation}
	\label{eq:Wf}
	f_j(k) = \varepsilon^{\ast\mu\nu}_j(k) F_{\mu\nu}(k)\,,
	\qquad
	F^{\mu\nu}(k) =\sum_{n}\frac{\sqrt{8\pi G}\, p_n^\mu p_n^\nu}{p_n\cdot k - i 0}
\end{equation}
and similarly for $f_{j}^{\text{out}/\text{in}}$ but restricting $n$ to final/initial states.  
We then define 
\begin{equation}
	\label{eikope4dbefore}
	\begin{split}
		&e^{2i\hat{\delta}(x_1,x_2)}   = \int\!\frac{d^D \tilde{Q}}{(2\pi)^D} \int\! d^D\tilde{x} \, e^{-i\tilde{Q}(\tilde{x}-x_1+x_2)} e^{i 2 \delta_s (b_e)}   \\
		&\times e^{\int_k \theta(\omega^\ast-k^0) 
			\left[
			f_j^\text{out} a_j^\dagger 
			- f_j^{\text{out}\ast} a_j
			\right]} 
		e^{\int_k \theta(\omega^\ast-k^0)\left[
			f_j^\text{in} a_j^\dagger 
			- f_j^{\text{in}\ast} a_j
			\right]}  \\
		&\times  e^{i\int_{k}\theta(k^0-\omega^\ast) \left[\tilde{\mathcal{A}}_j(x_1,x_2,k) a_j^\dagger(k)+
			\tilde{\mathcal{A}}_j^\ast(x_1,x_2,k) a_j(k)\right]}\,.
	\end{split}
\end{equation}
In contrast to \eqref{KF}, combining the two exponentials due to the dressings of initial and final states in the second line of \eqref{eikope4dbefore} \cite{Mirbabayi:2016axw,Choi:2017ylo,Addazi:2019mjh,Hannesdottir:2019opa} now produces a phase,
\begin{equation}\label{lastfactor}
	e^{\int_k^{\omega^\ast}\left[
		f_j^\text{out} a_j^\dagger 
		- f_j^{\text{out}\ast} a_j
		\right]}
	e^{\int_k^{\omega^\ast}\left[
		f_j^\text{in} a_j^\dagger 
		- f_j^{\text{in}\ast} a_j
		\right]}
	\simeq 
	e^{\int_k^{\omega^\ast}\left[
		f_j a_j^\dagger 
		- f_j^\ast a_j
		\right]} 
	e^{-2i\delta^\text{RR}}\,,
\end{equation}
which can be obtained performing the integrals as in \cite{DiVecchia:2022owy}. This crucially depends on the $-i0$ prescription, i.e.~on the fact that the dressings now include, unconventionally, static modes. This phase turns out to subtract the radiation-reaction part of the 3PM eikonal and leads us to define a ``conservative'' phase $2\tilde\delta$ via
\begin{equation}\label{Re2d2tilded}
	2\delta_s(b_e) 
	-
	2\tilde \delta(b_e)
	=
	 2\delta^\text{RR}(b_e)
	\simeq
	\tfrac14 G\, Q_\text{1PM}^2\,\mathcal I(\sigma) + \mathcal O(G^4)\,,
\end{equation}
with $\mathcal I(\sigma)$ and $Q_\text{1PM}$ as in Table~\ref{tab:omega=0} and Eq.~\eqref{Q1PM2PM} below. Eq.~\eqref{eikope4dbefore} then reads:
\begin{equation} \label{eikope4d}
	\begin{split}
          & e^{2i\hat{\delta}(x_1,x_2)}   = \int\!\frac{d^D \tilde{Q}}{(2\pi)^D} \int\! d^D\tilde{x} \, e^{-i\tilde{Q}(\tilde{x}-x_1+x_2)}  e^{i 2\tilde{\delta}(b_e)}  \\  & \times e^{\int_{k}\theta(\omega^\ast-k^0) \left[f_j a_j(k)^\dagger 
	- f_j^\ast(k) a_j(k)
	\right]}  e^{i\int_{k}\theta(k^0-\omega^\ast) \left[\tilde{\mathcal{A}}_j(x_1,x_2,k) a_j^\dagger(k)+\tilde{\mathcal{A}}_j^\ast(x_1,x_2,k) a_j(k)\right]}\,.
\end{split}
\end{equation}
In \eqref{eikope4d}, the initial momenta $p_{1,2}$ in $F^{\mu\nu}$ should be written in terms of the final momenta $p_{3,4}$ and of the $Q_i$. Note that the phase $2\tilde \delta$ still involves a ``democratic'' dependence between the incoming and the outgoing momenta via \eqref{deltael}, \eqref{Re2d2tilded}, so that Eq.~\eqref{eikope4d} again reproduces the classical Coulomb divergences due to the hard external particles~\cite{Weinberg:1965nx}.

The saddle-point conditions are then analogous to \eqref{Saddlev1} except for an extra contribution that arises from the static modes in the second line of~\eqref{eikope4d},
\begin{subequations}
	\label{Saddlev2}
\begin{align}
	\begin{split}
\label{Saddle1}
  (x_i - b_i)_\mu &=
  \frac{\partial 2\tilde\delta(b_e)}{\partial Q_i}
  -i\int_{k} \theta(\omega^\ast-k^0) \;F^\ast(k)\frac{\overset{\leftrightarrow}{\partial}}{\partial Q_i^\mu} F(k) \\ & \quad 
- i\int_{k}\theta(k^0-\omega^\ast)\; \tilde{\mathcal{A}}^\ast(x_1,x_2,k)\frac{\overset{\leftrightarrow}{\partial}}{\partial Q_i^\mu}\tilde{\mathcal{A}}(x_1,x_2,k)
 \,, 
 \end{split}
 \\ 
 \label{Saddle2}
Q_{i\,\mu} &=
(-1)^{i+1} \tilde{Q}_\mu -i\int_{k}\tilde{\mathcal{A}}^\ast(x_1,x_2,k)\frac{\overset{\leftrightarrow}{\partial}}{\partial x_i^\mu}\tilde{\mathcal{A}}(x_1,x_2,k)\;,~~
\\
\label{Saddle3}
\tilde{x}_\mu &= (x_1-x_2)_\mu +\frac{\partial 2\tilde\delta(b_e)}{\partial \tilde{Q}^\mu}
\,,
\qquad
\tilde{Q}_\mu = \frac{\partial 2\tilde\delta(b_e)}{\partial \tilde{x}^\mu}\,,
\end{align}
\end{subequations}
where in the second equation we took directly the limit $\omega^\ast\to 0$ since there are no static contributions.
We suppressed index contractions as in \eqref{suppressedindices}.

\section{3PM Observables in the Two-Body Problem}
\label{sec:observables}

We denote by $P^\alpha$ the energy-momentum of the gravitational field after the scattering and by $Q_{i}^\alpha$ the impulse for each particle, with
balance law $P^\alpha+Q^\alpha_{1}+Q^\alpha_{2}=0$.
We separate each observable into its radiative and non-radiative contributions. Radiative quantities arise from the part of the eikonal operator involving $\tilde{\mathcal{A}}^{\mu\nu}$, are due to graviton emissions with $\omega>\omega^\ast$ and will be denoted by bold symbols. For instance the only contribution to the momentum carried by the field is of this type, $P^\alpha=\boldsymbol{P}^\alpha$. The impulse of each particle contains also a non-radiative part, denoted with a subscript $(n)$, so that $Q_{i}^\alpha= \boldsymbol{Q}_{i}^\alpha +{Q}_{i(n)}^\alpha$.  
Following a similar notation we denote by $J^{\alpha\beta}$ the angular momentum of the gravitational field after the scattering and by $\Delta L_{i}^{\alpha\beta}$ the variation of angular momentum for each particle, whose balance law reads
$J^{\alpha\beta}
+
\Delta L_{1}^{\alpha\beta}
+
\Delta L_{2}^{\alpha\beta}
=
0$. 
In the following, we will show how all such quantities can be calculated from the eikonal operator to $\mathcal O(G^3)$ precision, providing in particular novel expressions for $\Delta L_{i}^{\alpha\beta}$.

As anticipated, our two proposals \eqref{eikope} and \eqref{eikope4d}  for  the eikonal operator lead to the same expressions for radiative quantities, while they differ in the way the non-radiative parts are treated.
In the first one, such non-radiative contributions arise from the phase $2\delta_s$ \eqref{deltael}. So, for the impulse, from \eqref{Saddlev1} we have
$
Q_{1(n)} = \frac{\partial 2\delta_s (b_e)}{\partial \tilde{x}}
$ 
as in \cite{DiVecchia:2021bdo}.
In the second one, one makes a finer distinction between conservative terms, which arise from the phase $2\tilde \delta$ and will be denoted with a subscript $(c)$, and static terms, which arise from $F^{\mu\nu}$ and  will be denoted by calligraphic symbols. For instance, from \eqref{Saddlev2}, $Q_{1(n)} = Q_{1(c)} + \mathcal{Q}_{1}$ with $Q_{1(c)} = \frac{\partial 2\tilde\delta(b_e)}{\partial \tilde{x}}$ as in \eqref{Qdd} and $\mathcal{Q}_{1}$ as in \eqref{RRQ1}. We will see that the final expression for $Q_{1(n)}$ up to 3PM order is the same in the two approaches.

When using \eqref{eikope}, one finds that ${J}^{\alpha\beta}$ is purely radiative, $J^{\alpha\beta}=\boldsymbol{J}^{\alpha\beta}$.
When using \eqref{eikope4d}, instead, it involves additional static-mode contributions, $J^{\alpha\beta} = \boldsymbol{J}^{\alpha\beta}+\mathcal J^{\alpha\beta}$. We will see that the latter expression for $J^{\alpha\beta}$ reproduces the results of~\cite{Damour:2020tta,Manohar:2022dea,DiVecchia:2022owy} for the gravitational angular momentum loss up to 3PM. 
This reflects the fact, emphasized in \cite{Veneziano:2022zwh}, that a standard $S$-matrix approach  necessarily leads to the canonical Bondi frame, hence to \eqref{eikope}, and thus misses the static-mode contributions captured by \eqref{eikope4d}.
Indeed, as discussed in \cite{Veneziano:2022zwh,DiVecchia:2022owy}, the difference between the two results for $J^{\alpha\beta}$, i.e.~the static contribution $\mathcal J^{\alpha\beta}$, can be seen as the effect of a BMS supertranslation (see \cite{Bonga:2018gzr} and references therein). The result $J^{\alpha\beta} = \boldsymbol{J}^{\alpha\beta}$ is the one obtained in the canonical Bondi frame, in which the initial Bondi shear vanishes, while the result $J^{\alpha\beta} = \boldsymbol{J}^{\alpha\beta}+\mathcal J^{\alpha\beta}$ is the one obtained in the ``intrinsic"  Bondi frame defined in \cite{Veneziano:2022zwh}, where the linear response formula of Bini and Damour \cite{Bini:2012ji,Damour:2020tta} should apply.
Similarly, using \eqref{eikope4d}, we will have both radiative, static and conservative contributions to the angular momentum changes, $\Delta L_{i}^{\alpha\beta}= \Delta \boldsymbol L_{i}^{\alpha\beta}+\Delta \mathcal L_{i} + \Delta L_{i\txt{(c)}}^{\alpha\beta}$. 

\subsection{Radiative contributions}
\label{sec:Radiative-modes} 

We first discuss the radiative contributions to the linear and angular momenta at ${\cal O}(G^3)$. Our strategy, based on reverse unitarity \cite{Anastasiou:2002yz,Anastasiou:2002qz,Anastasiou:2003yy,Herrmann:2021lqe,Herrmann:2021tct}, is to rewrite the formulae obtained from the eikonal operator in terms of integrals involving a three-particle cut in momentum space.\footnote{A priori, each observable is given by the exact expectation value of the corresponding self-adjoint operator in the final state. Equivalently, it can be obtained by integrating over phase space the relevant exact inclusive cross section weighted by an observable-specific factor \cite{DeTar:1971pmj,Kosower:2018adc}. For our coherent-state eikonal operators, the $\mathcal O(G^3)$ calculation simplifies to the three-particle phase space integrals described in the text, in spite of the fact that the production of an indefinite number of gravitons is implicitly taken into account. This is how  observables end up having a classical limit.} The resulting integrals reduce to suitable cuts of two-loop integrals, which one can calculate in the soft region with differential equations \cite{DiVecchia:2021bdo,Herrmann:2021tct}. 
The energy-momentum
$\boldsymbol P^\alpha$ \cite{Herrmann:2021lqe,Herrmann:2021tct} is given by
	\begin{equation}\label{Pformula}
		\boldsymbol{P}^\alpha = \int_{ k} \tilde{\mathcal{A}}\, k^\alpha \tilde{\mathcal{A}}^\ast\,.
	\end{equation}
We also consider, as a novel application of the reverse-unitarity method, \cite{Cristofoli:2021jas}
\begin{equation}\label{QRRformula}
	\boldsymbol Q_{i\alpha}
	= \operatorname{Im}\int_{ k} \frac{\partial\tilde{\mathcal{A}}}{\partial x_i^\alpha}\, \tilde{\mathcal{A}}^\ast\,.
\end{equation}
The main step consists in rewriting each of these observables $\boldsymbol O$ in the form
\begin{equation}\label{3pcqspacesimpl}
	\boldsymbol O
	=
	\operatorname{FT}
	\int
	d(\text{LIPS})
	f_{\boldsymbol{O}}
	\begin{gathered}
		\begin{tikzpicture}[scale=.5]
			\draw[<-] (-4.8,5.17)--(-4.2,5.17);
			\draw[<-] (-1,5.15)--(-1.6,5.15);
			\draw[<-] (-1,3.15)--(-1.6,3.15);
			\draw[<-] (-1,.85)--(-1.6,.85);
			\draw[<-] (-4.8,.83)--(-4.2,.83);
			\draw[<-] (-2.85,1.7)--(-2.85,2.4);
			\draw[<-] (-2.85,4.3)--(-2.85,3.6);
			\path [draw, thick, blue] (-5,5)--(-3,5)--(-1,5);
			\path [draw, thick, color=green!60!black] (-5,1)--(-3,1)--(-1,1);
			\path [draw] (-3,3)--(-1,3);
			\path [draw] (-3,1)--(-3,5);
			\draw[dashed] (-3,3) ellipse (1.3 and 2.3);
			\node at (-1,3)[below]{$k$};
			\node at (-5,5)[left]{$p_1$};
			\node at (-5,1)[left]{$p_2$};
			\node at (-2.8,4)[left]{$q_1$};
			\draw[<-] (3.35,5.17)--(2.75,5.17);
			\draw[<-] (-.45,5.15)--(.15,5.15);
			\draw[<-] (-.45,3.15)--(.15,3.15);
			\draw[<-] (-.45,.85)--(.15,.85);
			\draw[<-] (3.35,.83)--(2.75,.83);
			\draw[<-] (1.4,1.7)--(1.4,2.4);
			\draw[<-] (1.4,4.3)--(1.4,3.6);
			\path [draw, thick, red] (-.7,0)--(-.7,6);
			\path [draw, thick, blue] (3.55,5)--(1.55,5)--(-.45,5);
			\path [draw, thick, color=green!60!black] (3.55,1)--(1.55,1)--(-.45,1);
			\path [draw] (1.55,3)--(-.45,3);
			\path [draw] (1.55,1)--(1.55,5);
			\draw[dashed] (1.55,3) ellipse (1.3 and 2.3);
			\node at (1.35,4)[right]{$q-q_1$};
		\end{tikzpicture}
	\end{gathered}
\end{equation}
where each diagram represents $\mathcal{A}^{\mu\nu}$ with the appropriate routing and $d(\text{LIPS})$ stands for the
Lorentz-invariant phase space measure in the soft region,  
\begin{equation}\label{}
	\frac{d^Dk}{(2\pi)^D}2\pi\theta(k^0)\delta(k^2)
	\frac{d^Dq_1}{(2\pi)^D}2\pi\delta(2p_1\cdot q_1)
	2\pi\delta(2p_2\cdot (q_1 + k))\,.
\end{equation}
Note that $q^\alpha$ is the integrated variable in the Fourier transform $\mathrm{FT}[\,\cdots]$ defined by \eqref{FT2}.
The ``measuring function'' $f_{\boldsymbol{O}}$ corresponding to \eqref{Pformula} reads
$f_{\boldsymbol P^\alpha}=k^\alpha$ \cite{Herrmann:2021lqe,Herrmann:2021tct}. To obtain \eqref{QRRformula}, the appropriate function can be obtained by observing that, when the derivative acts on ${\tilde{\cal{A}}}$ in \eqref{A}, one gets a factor
$i q_1^\alpha$, while, when it acts on its complex conjugate, one gets $-i(q^\alpha-q_1^\alpha)$, so that $f_{\boldsymbol Q_{1}^\alpha}=q_1^\alpha-\frac12 q^\alpha$. The expression
$f_{\boldsymbol{Q}_{2}^\alpha}=-k^\alpha-q_1^\alpha+\frac12 q^\alpha$ can be obtained in an analogous way. In order to write the result in a compact form, we introduce the four-velocities $u_{i}^\mu$ for $i=1,2$, with 
	$u_{i}^\mu=-\frac{p_i^\mu}{m_i}$,
	$u_i^2=-1$,
	$\sigma = -u_1\cdot u_2$
and the variables \cite{Herrmann:2021tct}
\begin{equation}\label{check}
	\check u_1^\mu = \frac{\sigma \,u_2^\mu-u_1^\mu}{\sigma^2-1}\,,\qquad
	\check u_2^\mu = \frac{\sigma \,u_1^\mu-u_2^\mu}{\sigma^2-1}\,,
\end{equation}
which obey $\check u_i\cdot u_j =-\delta_{ij}$. In agreement with \cite{Herrmann:2021lqe,Herrmann:2021tct}, we obtain
\begin{equation}\label{Prad}
	\boldsymbol P^\alpha \simeq \frac{G^3m_1^2m_2^2}{b^3} \left(\check u_1^\mu+\check u_2^\mu\right)  \mathcal{E}
\end{equation}
with $\mathcal E$ as in Table~\ref{tab:omega>0} and $\check{u}_i$ as in \eqref{check}. 
\begin{table}[h]
	\fbox{
		\begin{minipage}{.96\linewidth}
			\begin{align*}
				\frac{\mathcal E}{\pi} 
				&= f_1 + f_2 \log\frac{\sigma+1}{2} + f_3 \frac{\sigma\,\operatorname{arccosh}\sigma}{2\sqrt{\sigma^2-1}}\\
				\frac{\mathcal C}{\pi} & = g_1 + g_2 \log\frac{\sigma+1}{2} + g_3 \frac{\sigma\,\operatorname{arccosh}\sigma}{2\sqrt{\sigma^2-1}} \\
				f_1 &= \frac{210\sigma^6-552\sigma^5+339\sigma^4-912\sigma^3+3148\sigma^2-3336\sigma+1151}{48(\sigma^2-1)^{3/2}}\\
				f_2 &= -\frac{35 \sigma^4+60 \sigma^3-150 \sigma^2+76 \sigma -5}{8\sqrt{\sigma^2-1}}\\
				f_3 &= \frac{(2 \sigma^2-3) \left(35 \sigma^4-30 \sigma^2+11\right)}{8 \left(\sigma^2-1\right)^{3/2}}\\
				g_1 &= \frac{105\sigma^7-411 \sigma^6+240\sigma^5+537\sigma^4-683\sigma^3+111\sigma^2+386\sigma-237}{24(\sigma^2-1)^{2}}\\
				g_2 &= \frac{35 \sigma ^5-90 \sigma ^4-70 \sigma ^3+16 \sigma ^2+155 \sigma -62}{4(\sigma^2-1)}\\
				g_3 &= -\frac{(2 \sigma ^2-3) \left(35 \sigma ^5-60 \sigma ^4-70 \sigma ^3+72 \sigma ^2+19 \sigma -12\right)}{4 \left(\sigma^2-1\right)^{2}}
			\end{align*}
		\end{minipage}
	}
	\caption{\label{tab:omega>0}
		Functions entering the radiative terms.}
\end{table} 
We also recover from \eqref{QRRformula} the longitudinal part of the RR impulse \cite{Herrmann:2021tct},
\begin{equation}\label{Q1Q2RR}
	\boldsymbol Q_{1}^{\alpha}
	\simeq
	- \frac{G^3m_1^2m_2^2}{b^3}\, \check{u}^\alpha_2\,\mathcal{E}
	\,,\qquad
	\boldsymbol Q_{2}^{\alpha}
	\simeq
	- \frac{G^3m_1^2m_2^2}{b^3}\, \check{u}^\alpha_1\,\mathcal{E}
	\,.
      \end{equation}
Since we are working at leading order (3PM) we used $p_4\simeq -p_1$ and $p_3\simeq -p_2$.  We observe that \eqref{Prad} and \eqref{Q1Q2RR} obey the balance law $\boldsymbol{P}^\alpha+\boldsymbol{Q}_{1}^\alpha+\boldsymbol{Q}_{2}^\alpha=0$.

In the framework based on \eqref{eikope}, the transverse part of the RR impulse, together with the conservative potential part, arises from the first term in~\eqref{saddle2}. As shown in \cite{DiVecchia:2021ndb} this can be related by analyticity to the zero-frequency limit of the graviton spectrum. In the alternative  formulation based on \eqref{eikope4d}, instead, the transverse part of the RR impulse will arise from the static contribution $\mathcal Q_{i}^\alpha$ \eqref{RRQ1}, while the conservative impulse is given by the derivative of $2\tilde\delta$ as in \eqref{Qdd} below.

The angular momentum carried away by gravitational waves in a $2\to2$ collision $\boldsymbol{J}^{\alpha\beta}$ was calculated to 3PM order in \cite{Manohar:2022dea} by resumming its small-velocity series. This was made possible by the assumption, checked a posteriori, that the resummed result be expressible using the same analytic functions appearing in the 3PM radiated energy-momentum $\boldsymbol{P}^\alpha$ \cite{Herrmann:2021lqe,Herrmann:2021tct}. Here we obtain again the result of \cite{Manohar:2022dea} for $\boldsymbol J^{\alpha\beta}$, using reverse unitarity. 
The formula for the radiated angular momenta in terms of $\tilde{\mathcal{A}}^{\mu\nu}$ is \cite{Manohar:2022dea,DiVecchia:2022owy} $\boldsymbol J_{\alpha\beta} = \boldsymbol J^{(o)}_{\alpha\beta} + \boldsymbol J^{(s)}_{\alpha\beta}$ with
\begin{equation}\label{Jgravexpl}
i\boldsymbol J^{(o)}_{\alpha\beta}
=
\int_{{k}}
k_{[\alpha}
\frac{\partial\tilde{\mathcal{A}}}{\partial k^{\beta]}}\tilde{\mathcal{A}}^\ast
\,,\qquad
\boldsymbol J^{(s)}_{\alpha\beta}
=
i \int_{{k}}
2\tilde{\mathcal{A}}^{\mu}_{[\alpha} \tilde{\mathcal{A}}_{\beta]\mu}^\ast\,.
\end{equation}
Similarly, inserting the appropriate differential operator $-ip_{4[\alpha}\frac{\partial}{\partial p_4^{\beta]}}$ in the eikonal operator as mentioned above, we derive the new formula $\Delta \boldsymbol L_{i}^{\alpha\beta} = \operatorname{Im}\boldsymbol{J}_{i}^{\alpha\beta}+b_i^{[\alpha}  \boldsymbol{Q}^{\beta]}_{i}$, where we define
\begin{equation}\label{DeltaLExpl}
\boldsymbol{J}_{i\alpha\beta}
=
\int_{{k}}
p_{i[\alpha}
\frac{\partial\tilde{\mathcal{A}}}{\partial p_i^{\beta]}}
\,
\tilde{\mathcal{A}}^\ast\,.
\end{equation}
These expressions start at ${\cal O}( G^3)$ and so, at our level of precision, we can use~\eqref{saddle1} and identify $x_j$ with $b_j$. Then it is convenient to use a translation to set $b_2=0$ and, to leading PM order, Eq.~\eqref{A} reduces to
\begin{equation}\label{Awhenb2=0}
	\tilde{\mathcal{A}}^{\mu\nu}(b,k) = \int \frac{d^Dq_1}{(2\pi)^D}\,2\pi\delta(2p_1\cdot q_1)\, e^{ib\cdot q_1} 2\pi\delta(2p_2\cdot (-q_1-k))\,\mathcal A^{\mu\nu}(q_1,-q_1-k,k)\,.
\end{equation}
The advantage of \eqref{Awhenb2=0} is that the phase factor and the first $\delta$-function are $k$-independent. At the end of the calculation we can perform a translation \eqref{translation} to another frame  by using
$\boldsymbol{J}^{\alpha\beta}\to \boldsymbol J^{\alpha\beta}+a^{[\alpha} \boldsymbol{P}^{\beta]}$ \cite{Manohar:2022dea} and we similarly obtain $\Delta\boldsymbol{L}_{i}^{\alpha\beta}\to \Delta\boldsymbol L_{i}^{\alpha\beta}+a^{[\alpha} \boldsymbol{Q}_{i}^{\beta]}$.\footnote{This holds even if $a^\mu= \rho_1b_1^\mu+\rho_2b_2^\mu$ and $\rho_{1,2}$ depend on $p_1$, $p_2$, because $\boldsymbol{P}\cdot b_{1,2}=0$.}

A key step analogous to \eqref{3pcqspacesimpl} in the calculation of angular momenta, starting from \eqref{Jgravexpl}, \eqref{DeltaLExpl},
seems more involved due to the presence of derivatives, which act in particular on the $\delta$ functions in \eqref{Awhenb2=0}.
The presence of $\delta'$ distributions may even seem to spoil the on-shell nature of the integration.
We resolve this difficulty as follows, focusing for simplicity on a frame where $b_2=0$. Although $\mathcal A^{\mu\nu}$ entering \eqref{Awhenb2=0} can be always modified by terms that vanish on-shell, we can choose a specific form for it and let the derivatives act both on $\mathcal A^{\mu\nu}$ and on the $\delta$ functions. Then, the resulting terms separately depend on the choice made at the beginning, but the final result is independent of it.
Proceeding in this way, and using the convenient distribution identity $\delta'(x)\delta(x-y)=\delta'(x)\delta(y)+\delta(x)\delta'(y)$ as an intermediate step, we find the following expressions suitable for the application of reverse unitarity:
\begin{equation}\label{LL1Lp}
	\begin{split}
		i\boldsymbol{J}^{(o)}_{\alpha\beta}
		&=
		\operatorname{FT}
		\int
	 k_{[\alpha}
		\frac{\partial }{\partial k^{\beta]}}
			\left[ d(\text{LIPS})
		\begin{gathered}
			\begin{tikzpicture}[scale=.5]
				\draw[<-] (-4.8,5.17)--(-4.2,5.17);
				\draw[<-] (-1,5.15)--(-1.6,5.15);
				\draw[<-] (-1,3.15)--(-1.6,3.15);
				\draw[<-] (-1,.85)--(-1.6,.85);
				\draw[<-] (-4.8,.83)--(-4.2,.83);
				\draw[<-] (-2.85,1.7)--(-2.85,2.4);
				\draw[<-] (-2.85,4.3)--(-2.85,3.6);
				\path [draw, thick, blue] (-5,5)--(-3,5)--(-1,5);
				\path [draw, thick, color=green!60!black] (-5,1)--(-3,1)--(-1,1);
				\path [draw] (-3,3)--(-1,3);
				\path [draw] (-3,1)--(-3,5);
				\draw[dashed] (-3,3) ellipse (1.3 and 2.3);
				\node at (-1,3)[below]{$k$};
				\node at (-5,5)[left]{$p_1$};
				\node at (-5,1)[left]{$p_2$};
				\node at (-2.8,4)[left]{$q_1$};
			\end{tikzpicture}
		\end{gathered}
		\right]
		\begin{gathered}
			\begin{tikzpicture}[scale=.5]
				\draw[<-] (4.8,5.17)--(4.2,5.17);
				\draw[<-] (1,5.15)--(1.6,5.15);
				\draw[<-] (1,3.15)--(1.6,3.15);
				\draw[<-] (1,.85)--(1.6,.85);
				\draw[<-] (4.8,.83)--(4.2,.83);
				\draw[<-] (2.85,1.7)--(2.85,2.4);
				\draw[<-] (2.85,4.3)--(2.85,3.6);
				\path [draw, thick, blue] (5,5)--(3,5)--(1,5);
				\path [draw, thick, color=green!60!black] (5,1)--(3,1)--(1,1);
				\path [draw] (3,3)--(1,3);
				\path [draw] (3,1)--(3,5);
				\draw[dashed] (3,3) ellipse (1.3 and 2.3);
				\node at (2.8,4)[right]{$q-q_1$};
			\end{tikzpicture}
		\end{gathered}
		\\
		&
		-
		u_{2[\alpha}
		\operatorname{FT} 
		\frac{\partial}{\partial q_{\parallel2}}
		\int d(\text{LIPS})
		k^{\phantom{2}}_{\beta]}
		\begin{gathered}
		\begin{tikzpicture}[scale=.5]
			\draw[<-] (-4.8,5.17)--(-4.2,5.17);
			\draw[<-] (-1,5.15)--(-1.6,5.15);
			\draw[<-] (-1,3.15)--(-1.6,3.15);
			\draw[<-] (-1,.85)--(-1.6,.85);
			\draw[<-] (-4.8,.83)--(-4.2,.83);
			\draw[<-] (-2.85,1.7)--(-2.85,2.4);
			\draw[<-] (-2.85,4.3)--(-2.85,3.6);
			\path [draw, thick, blue] (-5,5)--(-3,5)--(-1,5);
			\path [draw, thick, color=green!60!black] (-5,1)--(-3,1)--(-1,1);
			\path [draw] (-3,3)--(-1,3);
			\path [draw] (-3,1)--(-3,5);
			\draw[dashed] (-3,3) ellipse (1.3 and 2.3);
			\node at (-1,3)[below]{$k$};
			\node at (-5,5)[left]{$p_1$};
			\node at (-5,1)[left]{$p_2$};
			\node at (-2.8,4)[left]{$q_1$};
			\draw[<-] (3.35,5.17)--(2.75,5.17);
			\draw[<-] (-.45,5.15)--(.15,5.15);
			\draw[<-] (-.45,3.15)--(.15,3.15);
			\draw[<-] (-.45,.85)--(.15,.85);
			\draw[<-] (3.35,.83)--(2.75,.83);
			\draw[<-] (1.4,1.7)--(1.4,2.4);
			\draw[<-] (1.4,4.3)--(1.4,3.6);
			\path [draw, thick, red] (-.7,0)--(-.7,6);
			\path [draw, thick, blue] (3.55,5)--(1.55,5)--(-.45,5);
			\path [draw, thick, color=green!60!black] (3.55,1)--(1.55,1)--(-.45,1);
			\path [draw] (1.55,3)--(-.45,3);
			\path [draw] (1.55,1)--(1.55,5);
			\draw[dashed] (1.55,3) ellipse (1.3 and 2.3);
			\node at (1.35,4)[right]{$q-q_1$};
		\end{tikzpicture}
	\end{gathered}
	\end{split}
\end{equation}
where the derivative in the first line can act both on $\mathcal A^{\mu\nu}$ and on $d(\text{LIPS})$, 
and the derivative in the second line is with respect to the component $q_{\parallel2}$ of $q^\mu$ defined by
	$q^\mu = q_{\parallel1}\,\check u_1^\mu + q_{\parallel2} \,\check u_{2}^\mu + q_\perp^\mu$,
where $q_\perp \cdot u_{i}=0$. Indeed, while the Fourier transform \eqref{FT2} is eventually evaluated setting $q_{\parallel1}=q_{\parallel2}=0$, the derivative, arising from a $\delta'$, picks up the linear dependence on $q_{\parallel2}$ in the integrand.
For
$\boldsymbol {J}^{(s)}_{\alpha\beta}$, which involves no derivative,
one can apply \eqref{3pcqspacesimpl} with $f_{\boldsymbol{O}}=1$ and with the same index contractions as its $b$-space expression \eqref{Jgravexpl}.
Finally,
\begin{equation}\label{LJ2}
	\begin{split}
	\boldsymbol{J}_{2\alpha\beta}
		&=
\operatorname{FT}
\int
 u_{2[\alpha}
\frac{\partial }{\partial u_2^{\beta]}}
	\left[ d(\text{LIPS})
\begin{gathered}
	\begin{tikzpicture}[scale=.5]
		\draw[<-] (-4.8,5.17)--(-4.2,5.17);
		\draw[<-] (-1,5.15)--(-1.6,5.15);
		\draw[<-] (-1,3.15)--(-1.6,3.15);
		\draw[<-] (-1,.85)--(-1.6,.85);
		\draw[<-] (-4.8,.83)--(-4.2,.83);
		\draw[<-] (-2.85,1.7)--(-2.85,2.4);
		\draw[<-] (-2.85,4.3)--(-2.85,3.6);
		\path [draw, thick, blue] (-5,5)--(-3,5)--(-1,5);
		\path [draw, thick, color=green!60!black] (-5,1)--(-3,1)--(-1,1);
		\path [draw] (-3,3)--(-1,3);
		\path [draw] (-3,1)--(-3,5);
		\draw[dashed] (-3,3) ellipse (1.3 and 2.3);
		\node at (-1,3)[below]{$k$};
		\node at (-5,5)[left]{$p_1$};
		\node at (-5,1)[left]{$p_2$};
		\node at (-2.8,4)[left]{$q_1$};
	\end{tikzpicture}
\end{gathered}
\right]
\begin{gathered}
	\begin{tikzpicture}[scale=.5]
		\draw[<-] (4.8,5.17)--(4.2,5.17);
		\draw[<-] (1,5.15)--(1.6,5.15);
		\draw[<-] (1,3.15)--(1.6,3.15);
		\draw[<-] (1,.85)--(1.6,.85);
		\draw[<-] (4.8,.83)--(4.2,.83);
		\draw[<-] (2.85,1.7)--(2.85,2.4);
		\draw[<-] (2.85,4.3)--(2.85,3.6);
		\path [draw, thick, blue] (5,5)--(3,5)--(1,5);
		\path [draw, thick, color=green!60!black] (5,1)--(3,1)--(1,1);
		\path [draw] (3,3)--(1,3);
		\path [draw] (3,1)--(3,5);
		\draw[dashed] (3,3) ellipse (1.3 and 2.3);
		\node at (2.8,4)[right]{$q-q_1$};
	\end{tikzpicture}
\end{gathered}
\\
&	+
u_{2[\alpha}
\operatorname{FT} 
\frac{\partial}{\partial q_{\parallel2}}
\int  d(\text{LIPS})
(q_1+k)^{\phantom{2}}_{\beta]}
\begin{gathered}
	\begin{tikzpicture}[scale=.5]
		\draw[<-] (-4.8,5.17)--(-4.2,5.17);
		\draw[<-] (-1,5.15)--(-1.6,5.15);
		\draw[<-] (-1,3.15)--(-1.6,3.15);
		\draw[<-] (-1,.85)--(-1.6,.85);
		\draw[<-] (-4.8,.83)--(-4.2,.83);
		\draw[<-] (-2.85,1.7)--(-2.85,2.4);
		\draw[<-] (-2.85,4.3)--(-2.85,3.6);
		\path [draw, thick, blue] (-5,5)--(-3,5)--(-1,5);
		\path [draw, thick, color=green!60!black] (-5,1)--(-3,1)--(-1,1);
		\path [draw] (-3,3)--(-1,3);
		\path [draw] (-3,1)--(-3,5);
		\draw[dashed] (-3,3) ellipse (1.3 and 2.3);
		\node at (-1,3)[below]{$k$};
		\node at (-5,5)[left]{$p_1$};
		\node at (-5,1)[left]{$p_2$};
		\node at (-2.8,4)[left]{$q_1$};
		\draw[<-] (3.35,5.17)--(2.75,5.17);
		\draw[<-] (-.45,5.15)--(.15,5.15);
		\draw[<-] (-.45,3.15)--(.15,3.15);
		\draw[<-] (-.45,.85)--(.15,.85);
		\draw[<-] (3.35,.83)--(2.75,.83);
		\draw[<-] (1.4,1.7)--(1.4,2.4);
		\draw[<-] (1.4,4.3)--(1.4,3.6);
		\path [draw, thick, red] (-.7,0)--(-.7,6);
		\path [draw, thick, blue] (3.55,5)--(1.55,5)--(-.45,5);
		\path [draw, thick, color=green!60!black] (3.55,1)--(1.55,1)--(-.45,1);
		\path [draw] (1.55,3)--(-.45,3);
		\path [draw] (1.55,1)--(1.55,5);
		\draw[dashed] (1.55,3) ellipse (1.3 and 2.3);
		\node at (1.35,4)[right]{$q-q_1$};
	\end{tikzpicture}
\end{gathered}
\end{split}
\end{equation}
for the integral entering \eqref{DeltaLExpl} for $i=2$. 

Although we discussed the above steps for $b_2=0$, we present the final results in a frame where $b_1+b_2=0$, related to the previous one by a translation by $-b/2$, where particle-interchange symmetry is manifest.
Defining $\mathcal E_{\pm}$ and $\mathcal F$ in terms of the functions $\mathcal E$, $\mathcal C$ given in Table~\ref{tab:omega>0}  \cite{Herrmann:2021lqe,Herrmann:2021tct,Manohar:2022dea}
via $\mathcal C {\sqrt{\sigma^2-1}} = -\mathcal E_+ +\sigma \mathcal E_-$ and $\mathcal F=\mathcal E_+-\tfrac12\,\mathcal E= - \mathcal E_- +\tfrac12 \mathcal E$,
 we find
\begin{equation}\label{JabF}
	\boldsymbol{J}^{\alpha\beta}\simeq\frac{G^3m_1^2m_2^2}{b^3} \,\mathcal F \left(
	b^{[\alpha}\check u_{1}^{\beta]}
	-
	b^{[\alpha}\check u_{2}^{\beta]}
	\right).
\end{equation}
After a translation with $a^\mu=\frac{E_2 - E_1}{2(E_1+E_2)} b^\mu$, Eq.~\eqref{JabF} reproduces Eq.~(15) of Ref.~\cite{Manohar:2022dea} \emph{except} for the static (zero-frequency) modes. 
We shall see below how those terms can be recovered in the present formalism.
Moreover, we obtain the new result
\begin{equation}\label{deltabL1}
	\Delta  \boldsymbol L_{1}^{\alpha \beta}\simeq \frac{G^{3} m_{1}^{2} m_{2}^{2}}{b^{3}}\left[\frac{\mathcal{E}_{+} b^{[\alpha}u_{1}^{\beta]}}{\sigma-1} -\frac{1}{2}\,\mathcal{E}\, b^{[\alpha} \check{u}_{2}^{\beta]}\right]
\end{equation}
($\Delta  \boldsymbol L_{2}$ is obtained by $1\leftrightarrow 2$ and $b^\alpha\leftrightarrow -b^\alpha$). The radiative quantities obey the balance law
$\boldsymbol J^{\alpha\beta}
+
\Delta \boldsymbol L_{1}^{\alpha\beta}
+
\Delta \boldsymbol L_{2}^{\alpha\beta}
=
0$. 

We conclude by calculating the integral needed to evaluate the second line of \eqref{Saddle1}.
To this end, let us define the radiative contribution to $x_i-b_i$,
\begin{equation}\label{}
	\delta x_{i\mu}
	=
	- i\int_{k} \tilde{\mathcal{A}}^\ast(x_1,x_2,k)\frac{\overset{\leftrightarrow}{\partial}}{\partial Q_i^\mu}\tilde{\mathcal{A}}(x_1,x_2,k)
	 \,.
\end{equation}
Noting that $\partial_{Q_i}\tilde{\mathcal{A}}=-\tfrac12\partial_{p_i}\tilde{\mathcal{A}}$, we see that the integrals needed to evaluate $\tilde Q\cdot \delta x_i$ are the same as those entering the calculation of $\Delta \boldsymbol L_{i}^{\alpha\beta}$, and we obtain for $\delta x_1-\delta x_2=\delta x$
\begin{equation}\label{QRRdynamicAdA}
	\tilde{Q} \cdot \delta x \simeq \frac{m_1+m_2}{2} \frac{G^3 \tilde{Q} m_1 m_2}{\left(\sigma-1\right) b^2} \mathcal{E}_+(\sigma)\,.
\end{equation} 

\subsection{Static contributions}
\label{sec:statcont}

The angular momentum of the time-independent field was calculated to 2PM order in \cite{Damour:2020tta} and to 3PM order in \cite{Manohar:2022dea,DiVecchia:2022owy}. In the present approach it is natural to start from the formulation in~\eqref{eikope4d} for the eikonal operator to derive the static contribution to the angular momentum of the gravitational field. By following~\cite{DiVecchia:2022owy} one obtains 
\begin{equation}\label{calJab}
	\mathcal J_{\alpha\beta} = -i\int_{ k}\left( F^\ast\,k_{[\alpha}\frac{\partial F}{\partial k^{\beta]}}+2F^{\ast}_{\mu[\alpha}F^\mu_{\beta]}\right).
\end{equation}
For the purposes of this work we use the 2PM approximation to describe the gravitational collision and at this order we have $Q_1\simeq -Q_2 \simeq Q \simeq \tilde{Q}$. 
In terms of the coefficients $c_{nm}$ defined in Table~\ref{tab:omega=0}, 
for a $2\to2$ collision
the result of \eqref{calJab}  
reads (see \cite[Eq.~(3.30)]{DiVecchia:2022owy} for a more detailed discussion)
\begin{equation}\label{Jabnm}
	\mathcal J^{\alpha\beta} = - \sum_{n=1,2}\sum_{m=3,4} c_{nm}\,p_n^{[\alpha} p_m^{\beta]}\,.
\end{equation}
To this order $\sigma_{12}=\sigma\simeq\sigma_{34}$, 
	$\sigma_{13}\simeq\sigma_{24}\simeq\sigma-\frac{Q^{2}}{2 m_{1} m_{2}}$,
	$\sigma_{14}\simeq 1+\frac{Q^{2}}{2 m_{1}^{2}}$,  
	$\sigma_{23} \simeq 1+\frac{Q^{2}}{2 m_{2}^{2}}$.
\begin{table}[h]
	\fbox{
		\begin{minipage}{.65\linewidth}
			\begin{align*}
				\sigma_{nm}&=-\eta_n\eta_m\,\frac{p_n\cdot p_m}{m_n m_m}
				\,,
				\\
				c_{nm} &= 
				2G\left[ \left(
				\sigma_{nm}^2-\tfrac12
				\right) \frac{\sigma_{nm}\Delta_{nm}-1}{\sigma_{nm}^2-1}
				-
				2
				\sigma_{nm}\Delta_{nm}
				\right],
				\\
				\tfrac12\,\mathcal{I}&=\frac{8-5\sigma^2}{3(\sigma^2-1)}
				+
				\frac{\sigma(2\sigma^2-3)\operatorname{arccosh}\sigma}{(\sigma^2-1)^{3/2}}\,.
			\end{align*}
		\end{minipage}%
		\begin{minipage}{.3\linewidth}
			\vspace{-12pt}
			\begin{align*}
				\Delta_{nm}&=\frac{\operatorname{arccosh}\sigma_{nm}}{\sqrt{\sigma^2_{nm}-1}}\,,
				\\
				d_{nm} &=2G\,\frac{\sigma^2_{nm}-\tfrac12}{\sigma_{nm}^2-1} \,,
				\\
				2\mathcal G
				&=
				c_{14}
				+
				c_{23}
				-
				2c_{24}
				\,,
			\end{align*}
		\end{minipage}%
	}
	\caption{\label{tab:omega=0}%
		Functions and coefficients entering the static terms. $\eta_n=+1$ ($\eta_n=-1$)  if the $n$th state is outgoing (incoming).}
\end{table} 

Expanding Eq.~\eqref{Jabnm} to its leading $\mathcal O(G^2)$ order using $Q \simeq Q_\text{1PM}$ as in \eqref{Q1PM2PM} reproduces the 2PM angular momentum loss \cite{Damour:2020tta}.
Summing the radiative contribution $\boldsymbol{J}^{\alpha\beta}$ \eqref{JabF} and the 3PM expansion of the static contribution $\mathcal{J}^{\alpha\beta}$ \eqref{Jabnm}, using $Q \simeq Q_\text{1PM}+Q_\text{2PM}$ as in \eqref{Q1PM2PM}, reproduces the full 3PM result of \cite{Manohar:2022dea}.

A similar technique allows one to calculate the portion of the RR impulse that is now due to zero-frequency modes.
The first step is to evaluate the integral in the first line of \eqref{Saddle1}.
To this end, let us isolate its contribution to $x-b$,
\begin{equation}\label{}
	 \Delta x_\mu
	=
	- i\int_{k}\theta(\omega^\ast-k^0)\; F^\ast(k)\Big(\frac{\overset{\leftrightarrow}{\partial}}{\partial Q_1^\mu}
	-
	\frac{\overset{\leftrightarrow}{\partial}}{\partial Q_2^\mu}\Big) F(k)
	\,.
\end{equation}
To the order under consideration,
using the on-shell conditions and the same integrals involved in the evaluation of $\mathcal{J}^{\alpha\beta}$, one obtains
\begin{equation}\label{QRRstaticFdF}
	\Delta x_\mu
	\simeq
	Q_\mu\,\mathcal G\,.
\end{equation}
This term contributes to the relation between $b_e$ and $b$ allowing one to obtain the impulse as a function of the impact parameter, via the second saddle point condition \eqref{Saddle3}.
To leading order, where ${\cal{G}}_{RR} \simeq \frac{G}{2} {\cal{I}}(\sigma)$, this produces the contribution
\begin{equation}\label{RRQ1}
	\mathcal Q^{\alpha}_{1}
	= 
	-\frac{b^\alpha}{2b^2}\,G\,Q^2_\text{1PM}\,\mathcal I(\sigma)+\mathcal O(G^4)
\end{equation}
and thus provides the 3PM RR corrections to the deflection angle \cite{Damour:2020tta}, so that \eqref{Q1Q2RR} and \eqref{RRQ1} reproduce the full RR impulse to 3PM order \cite{Herrmann:2021tct}. This enters the balance law in a trivial way, since 
$\mathcal Q_{1}=-\mathcal Q_{2}$.

Finally, we derive the following formula for the change in angular momentum of each particle due to the interaction with the static field, valid up to $\mathcal O(G^3)$,
\begin{equation}\label{DcalL}
	\begin{split}
	\Delta \mathcal L_{1}^{\alpha\beta} =\operatorname{Im}
	\int_{{k}}
		F^\ast
		\left(
	p_{4[\alpha}
	\frac{\partial}{\partial p_4^{\beta]}} 
	+
	Q_{1[\alpha}
	\frac{\partial}{\partial Q_1^{\beta]}} 
	\right) F
	+
b_1^{[\alpha}\mathcal{Q}^{\beta]}_{1}\,.
\end{split}
\end{equation} 
Under a translation \eqref{translation}, $\Delta \mathcal L_{i}^{\alpha\beta}\to \Delta \mathcal L_{i}^{\alpha\beta}+a_{\phantom{i}}^{[\alpha}\mathcal Q_{i}^{\beta]}$.
Defining for $m=1,2,3,4$,
\begin{equation}\label{}
	2\eta_m J_{(m)}^{\alpha\beta}
	=
	\sum_{\eta_n=-\eta_m} 
	c_{nm}\,
	p_n^{[\alpha}p_m^{\beta]} 
	-
	\sum_{\substack{\eta_n=\eta_m\\n\neq m}}d_{nm} \, p_n^{[\alpha}p_m^{\beta]}\,,
\end{equation}
we find the following result for \eqref{DcalL}, 
\begin{equation}\label{}
	\begin{split}
			\Delta \mathcal L_{1}^{\alpha\beta} =
		J_{1}^{\alpha\beta}
		+
		J_{4}^{\alpha\beta}
		+
		b_{1}^{[\alpha} \mathcal Q^{\beta]}_{1}\,.
	\end{split}
\end{equation}
These static contributions obey the balance law
$\mathcal J^{\alpha\beta}
+
\Delta \mathcal L_{1}^{\alpha\beta}
+
\Delta \mathcal L_{2}^{\alpha\beta}
=
0$.

\subsection{Conservative effects}

We conclude with a discussion of changes in the observables due to purely conservative effects, based on \eqref{eikope4d}. The conservative impulse is obtained from the second equation in \eqref{Saddle3} and reads 
\begin{equation}\label{Qdd}
	\tilde Q^\alpha= -\tilde Q\,\frac{b_{e}^{\alpha}}{b_e}\,,\qquad
	\tilde Q =- \frac{\partial\, 2\tilde\delta(b_e)}{\partial b_e}\,.
\end{equation}
For instance, using \eqref{delta1-2PM}, one finds
\begin{equation}\label{Q1PM2PM}
	Q_\text{1PM}
	=
	\frac{4G m_1m_2}{b_e}\,\frac{\sigma^2-\frac{1}{2}}{\sqrt{\sigma^2-1}}\,,
	\qquad
	Q_\text{2PM}
	=
	\frac{3\pi G^2 m_1m_2(m_1+m_2)}{4b^2_e}\,\frac{5\sigma^2-1}{\sqrt{\sigma^2-1}}\,.
\end{equation}   
Note that the Coulombic divergence in $2\delta_0$ does not affect ${Q}_\text{1PM}$. In the present setup, contrary to \cite{DiVecchia:2021bdo}, the transverse RR impulse at 3PM comes from \eqref{RRQ1} and not from the contribution of $2\tilde\delta_2$ to \eqref{Qdd}, due to its definition \eqref{Re2d2tilded}.

For the changes $\Delta L_{1(c)}^{\alpha\beta}$, we find
\begin{equation}\label{DeltaL1c}
	\begin{split}
\Delta L_{1(c)}^{\alpha\beta}
=
(b_{1}
-A_{12}\,
p_{1})^{[\alpha}_{\phantom{1}}\tilde Q^{\beta]}+
\left[
p_{1}^{[\alpha}p_{2}^{\beta]}+
\left(p_{1}+p_{2}\right)^{[\alpha} \tilde Q^{\beta]}
\right]
S
	\end{split}
\end{equation}
with
\begin{equation}\label{}
	A_{12} = \frac{m_2(m_1\sigma+m_2)}{2m_1^2m_2^2(\sigma^2-1)}\,\tilde Q\, b_e\,,\qquad S = \frac{\partial 2\tilde\delta(s, b_e)}{\partial(p_1\cdot p_2)}\,.
\end{equation}
The balance law $\Delta L_{1(c)}^{\alpha\beta}+\Delta L_{2(c)}^{\alpha\beta}=0$ holds in any reference frame and follows from the relation
\begin{equation}\label{tildebeb}
	b_e^\mu = b^\mu - A_{12}\, p_1^\mu + A_{21}\, p_2^{\mu}\,.
\end{equation}
Unlike $\tilde Q$, however, $S$ is sensitive to the cutoff $b_0$ in the subtraction of the Coulombic divergence.\footnote{An asymmetric choice of cutoff when subtracting this divergence would lead to an $\mathcal O(G)$ scoot term as in Refs.~\cite{Gralla:2021eoi,Gralla:2021qaf}.} This reflects the infinite Shapiro time-delay for scattering events in $D=4$ and makes the second term in \eqref{DeltaL1c} ambiguous. Yet, going to a frame where, initially, the center-of-mass is at rest $-p_1^\alpha=(E_1,p^I)$, $-p_2^\alpha=(E_2,-p^I)$ and sits in the origin in the transverse plane (C.o.M. frame)
	$b_1^\alpha=\frac{E_2}{E_1+E_2}\,b^\alpha$, 
	$b_2^\alpha=-\frac{E_1}{E_1+E_2}\,b^\alpha$,
we obtain that this ambiguity only appears in the variation of each particle's mass-dipole, $\Delta L_{i(c)}^{0I}$, a notoriously ill defined quantity \cite{Gralla:2021eoi,Gralla:2021qaf}. The variation of each particle's spatial angular momentum, $\Delta L_{i(c)}^{IJ}$, is instead perfectly well defined and in fact vanishes:
	$\Delta L_{i(c)}^{IJ}=0$,
as expected for the conservative process as seen from the C.o.M. frame.

\section{Outlook}
\label{sec:outlook}

In this letter we showed how the eikonal operator~\cite{Cristofoli:2021jas,DiVecchia:2022owy,DiVecchia:2022nna} can be used to derive explicit expressions for the 3PM classical observables in the gravitational scattering of two massive scalars. 
Of course it is also interesting to extend this approach to the 4PM order. In Section~\ref{sec:statcont} we saw that, in the formulation of the eikonal operator presented in Section~\ref{sec:form2}, the 3PM RR corrections~\eqref{RRQ1} naturally follow from 2PM data, exactly as in the purely classical linear response formula~\cite{Bini:2012ji,Damour:2020tta}. It is not difficult to see that a similar pattern holds also at the next order: the 3PM data we obtained can be used to find the 4PM linear response contribution to the transverse impulse~\cite{Manohar:2022dea} in~\eqref{Saddle2}. As a first step, we have to express the result in~\eqref{Saddle3} in terms of the initial data and, for instance, use
  $\sigma_{34} = \sigma_{12} - \tfrac{1}{m_1 m_2}\left[(p_1+p_2)\cdot P +\tfrac12 P^2\right]$
to rewrite $2\tilde{\delta}_0(\sigma_{34},b_e)$. This yields a new ${\cal O}(G^4)$ term involving the derivative of the 1PM eikonal phase times the ${\cal O}(G^3)$ radiated energy reproducing exactly a first term in the linear response formula of~\cite{Bini:2012ji}. Then, when the radiative ${\cal O}(G^3)$ part~\eqref{QRRdynamicAdA} in the relation between $b_e$ and $b$ is taken into account in~\eqref{Saddle3}, one gets a new linear response contribution.\footnote{This mechanism was discussed in~\cite{Cristofoli:2021jas} but focusing just on the conservative contributions.} By combining these two contributions with that from the static modes, which follows from~\eqref{QRRstaticFdF}, we get the ${\cal O}(G^4)$ radiation-reaction part of the impulse along the direction $b_e^\mu$
\begin{align}
  \label{eq:rrg4}
 Q_{1}^\text{RR} \simeq - Q_{2}^\text{RR} & \simeq  \left[ {\mathcal E} \frac{d}{d\sigma} \frac{2\sigma^2-1}{\sqrt{\sigma^2-1}}  +   \frac{(2\sigma^2-1)\sigma}{(\sigma^2-1)^{\frac{3}{2}}} {\mathcal E} - \frac{2\sigma^2-1}{\sigma^2-1} {\mathcal C} \right. \\ &\left. \quad - \frac{(2\sigma^2-1)3\pi (5\sigma^2-1)}{\sigma^2-1}  \frac{3{\mathcal I}}{4}\right] \frac{G^4 m^2_1 m^2_2 (m_1+m_2)}{b^4}\,, \nonumber 
\end{align}
where the first and second lines contain the radiative and the static parts respectively. In order to compare with~\cite{Bini:2021gat,Manohar:2022dea}, we should consider the projection of the impulse along $b^\mu$ instead of $b_e^\mu$. Then we need to ``undo'' the approximation done in~\eqref{Q1Q2RR} and use~\eqref{tildep}, which implies that the spatial part of $\boldsymbol Q_{1}$ is along $\vec{p}_1-\vec{p}_4$~\cite{Bini:2021gat}. Thus, at leading order, the results in~\eqref{Q1Q2RR} are orthogonal to $b_e$, but have a non-trivial  projection along $b$. This yields a new ${\cal O}(G^4)$ to the RR impulse
along $b^\mu$: $\sin\frac{\Theta_s}{2} |\vec{\boldsymbol Q}_{1}|$. By adding this new term to~\eqref{eq:rrg4}, we reproduce Eq.~(19) of~\cite{Manohar:2022dea}, which was recently confirmed \cite{Dlapa:2022lmu} and extended both in the PN \cite{Bini:2022enm} and the PM approach \cite{Dlapa:2022lmu}.

It would be interesting to compare our results with those obtained in~\cite{toap2} by using the KMOC approach~\cite{Kosower:2018adc} and also with Post-Newtonian data for the mechanical angular momenta. Testing the present formalism beyond what we have discussed so far, e.g.~including the 4PM longitudinal impulses and calculating the NLO corrections to the waveforms, will also shed light on whether the operator exponent $2\hat{\delta}$ needs non-linear terms in the creation/annihilation operators. For instance, we expect that terms of the type $a^\dagger a$, involving the $2\to 2$ Compton-like amplitude~\cite{KoemansCollado:2019ggb}, are needed.\footnote{In~\cite{Britto:2021pud} the 6-point tree amplitude with two gravitons was analysed and it was shown that it does not yield classical contributions to be included in the eikonal operator.}
  
\subsection*{Acknowledgements} 
We would like to thank Donato Bini, Thibault Damour, Riccardo Gonzo, Enrico Herrmann, Gregor K\"alin, Donal O'Connell, Julio Parra-Martinez, Rafael Porto, Michael Ruf, Mao Zeng for several enlightening discussions.
RR and GV are grateful to IHES for hospitality during the first part of this work.
CH and RR are grateful to KITP, Santa Barbara, for hospitality during the workshop High-Precision Gravitational Waves. 

This research was supported in part by the National Science Foundation under Grant No.~NSF PHY-1748958.
The research of RR is partially supported by the UK Science and Technology Facilities Council (STFC) Consolidated Grant ST/T000686/1. The research of CH (PDV) is fully (partially) supported by the Knut and Alice Wallenberg Foundation under grant KAW 2018.0116. Nordita is partially supported by Nordforsk.


\begin{thebibliography}{10}
	
	\bibitem{Manohar:2022dea}
	A.~V. Manohar, A.~K. Ridgway, and C.-H. Shen, ``{Radiated Angular Momentum and
		Dissipative Effects in Classical Scattering},''
	\href{http://dx.doi.org/10.1103/PhysRevLett.129.121601}{{\em Phys. Rev.
			Lett.} {\bf 129} (2022) no.~12, 121601},
	\href{http://arxiv.org/abs/2203.04283}{{\tt arXiv:2203.04283 [hep-th]}}.
	
	\bibitem{Amati:1987wq}
	D.~Amati, M.~Ciafaloni, and G.~Veneziano, ``{Superstring Collisions at
		Planckian Energies},''
	\href{http://dx.doi.org/10.1016/0370-2693(87)90346-7}{{\em Phys. Lett.} {\bf
			B197} (1987)  81}.
	
	\bibitem{Amati:1990xe}
	D.~Amati, M.~Ciafaloni, and G.~Veneziano, ``{Higher Order Gravitational
		Deflection and Soft Bremsstrahlung in Planckian Energy Superstring
		Collisions},''
	\href{http://dx.doi.org/10.1016/0550-3213(90)90375-N}{{\em Nucl. Phys.} {\bf
			B347} (1990)  550--580}.
	
	\bibitem{Kabat:1992tb}
	D.~N. Kabat and M.~Ortiz, ``{Eikonal quantum gravity and Planckian
		scattering},'' \href{http://dx.doi.org/10.1016/0550-3213(92)90627-N}{{\em
			Nucl. Phys.} {\bf B388} (1992)  570--592},
	\href{http://arxiv.org/abs/hep-th/9203082}{{\tt arXiv:hep-th/9203082
			[hep-th]}}.
	
	\bibitem{Akhoury:2013yua}
	R.~Akhoury, R.~Saotome, and G.~Sterman, ``{High Energy Scattering in
		Perturbative Quantum Gravity at Next to Leading Power},''
	\href{http://dx.doi.org/10.1103/PhysRevD.103.064036}{{\em Phys. Rev. D} {\bf
			103} (2021) no.~6, 064036}, \href{http://arxiv.org/abs/1308.5204}{{\tt
			arXiv:1308.5204 [hep-th]}}.
	
	\bibitem{Bjerrum-Bohr:2018xdl}
	N.~E.~J. Bjerrum-Bohr, P.~H. Damgaard, G.~Festuccia, L.~Plant{\'e}, and
	P.~Vanhove, ``{General Relativity from Scattering Amplitudes},''
	\href{http://dx.doi.org/10.1103/PhysRevLett.121.171601}{{\em Phys. Rev.
			Lett.} {\bf 121} (2018) no.~17, 171601},
	\href{http://arxiv.org/abs/1806.04920}{{\tt arXiv:1806.04920 [hep-th]}}.
	
	\bibitem{Ciafaloni:2015xsr}
	M.~Ciafaloni, D.~Colferai, F.~Coradeschi, and G.~Veneziano, ``{Unified limiting
		form of graviton radiation at extreme energies},''
	\href{http://dx.doi.org/10.1103/PhysRevD.93.044052}{{\em Phys. Rev.} {\bf
			D93} (2016) no.~4, 044052},
	\href{http://arxiv.org/abs/1512.00281}{{\tt arXiv:1512.00281 [hep-th]}}.
	
	\bibitem{Ciafaloni:2018uwe}
	M.~Ciafaloni, D.~Colferai, and G.~Veneziano, ``{Infrared features of
		gravitational scattering and radiation in the eikonal approach},''
	\href{http://dx.doi.org/10.1103/PhysRevD.99.066008}{{\em Phys. Rev.} {\bf
			D99} (2019) no.~6, 066008},
	\href{http://arxiv.org/abs/1812.08137}{{\tt arXiv:1812.08137 [hep-th]}}.
	
	\bibitem{Cristofoli:2021jas}
	A.~Cristofoli, R.~Gonzo, N.~Moynihan, D.~O'Connell, A.~Ross, M.~Sergola, and
	C.~D. White, ``{The Uncertainty Principle and Classical Amplitudes},''
	\href{http://arxiv.org/abs/2112.07556}{{\tt arXiv:2112.07556 [hep-th]}}.
	
	\bibitem{Addazi:2019mjh}
	A.~Addazi, M.~Bianchi, and G.~Veneziano, ``{Soft gravitational radiation from
		ultra-relativistic collisions at sub- and sub-sub-leading order},''
	\href{http://dx.doi.org/10.1007/JHEP05(2019)050}{{\em JHEP} {\bf 05} (2019)
		050},
	\href{http://arxiv.org/abs/1901.10986}{{\tt arXiv:1901.10986 [hep-th]}}.
	
	\bibitem{Damgaard:2021ipf}
	P.~H. Damgaard, L.~Plante, and P.~Vanhove, ``{On an exponential representation
		of the gravitational S-matrix},''
	\href{http://dx.doi.org/10.1007/JHEP11(2021)213}{{\em JHEP} {\bf 11} (2021)
		213}, \href{http://arxiv.org/abs/2107.12891}{{\tt arXiv:2107.12891
			[hep-th]}}.
	
	\bibitem{DiVecchia:2022owy}
	P.~Di~Vecchia, C.~Heissenberg, and R.~Russo, ``{Angular momentum of
		zero-frequency gravitons},''
	\href{http://dx.doi.org/10.1007/JHEP08(2022)172}{{\em JHEP} {\bf 08} (2022)
		172}, \href{http://arxiv.org/abs/2203.11915}{{\tt arXiv:2203.11915
			[hep-th]}}.
	
	\bibitem{DiVecchia:2022nna}
	P.~Di~Vecchia, C.~Heissenberg, R.~Russo, and G.~Veneziano, ``{The eikonal
		operator at arbitrary velocities I: the soft-radiation limit},''
	\href{http://dx.doi.org/10.1007/JHEP07(2022)039}{{\em JHEP} {\bf 07} (2022)
		039}, \href{http://arxiv.org/abs/2204.02378}{{\tt arXiv:2204.02378
			[hep-th]}}.
	
	\bibitem{Bern:2019nnu}
	Z.~Bern, C.~Cheung, R.~Roiban, C.-H. Shen, M.~P. Solon, and M.~Zeng,
	``{Scattering Amplitudes and the Conservative Hamiltonian for Binary Systems
		at Third Post-Minkowskian Order},''
	\href{http://dx.doi.org/10.1103/PhysRevLett.122.201603}{{\em Phys. Rev.
			Lett.} {\bf 122} (2019) no.~20, 201603},
	\href{http://arxiv.org/abs/1901.04424}{{\tt arXiv:1901.04424 [hep-th]}}.
	
	\bibitem{Bern:2019crd}
	Z.~Bern, C.~Cheung, R.~Roiban, C.-H. Shen, M.~P. Solon, and M.~Zeng, ``{Black
		Hole Binary Dynamics from the Double Copy and Effective Theory},''
	\href{http://dx.doi.org/10.1007/JHEP10(2019)206}{{\em JHEP} {\bf 10} (2019)
		206}, \href{http://arxiv.org/abs/1908.01493}{{\tt arXiv:1908.01493
			[hep-th]}}.
	
	\bibitem{Kalin:2020fhe}
	G.~K\"alin, Z.~Liu, and R.~A. Porto, ``{Conservative Dynamics of Binary Systems
		to Third Post-Minkowskian Order from the Effective Field Theory Approach},''
	\href{http://dx.doi.org/10.1103/PhysRevLett.125.261103}{{\em Phys. Rev.
			Lett.} {\bf 125} (2020) no.~26, 261103},
	\href{http://arxiv.org/abs/2007.04977}{{\tt arXiv:2007.04977 [hep-th]}}.
	
	\bibitem{Parra-Martinez:2020dzs}
	J.~Parra-Martinez, M.~S. Ruf, and M.~Zeng, ``{Extremal black hole scattering at
		$\mathcal{O}(G^3)$: graviton dominance, eikonal exponentiation, and
		differential equations},''
	\href{http://dx.doi.org/10.1007/JHEP11(2020)023}{{\em JHEP} {\bf 11} (2020)
		023}, \href{http://arxiv.org/abs/2005.04236}{{\tt arXiv:2005.04236
			[hep-th]}}.
	
	\bibitem{DiVecchia:2020ymx}
	P.~Di~Vecchia, C.~Heissenberg, R.~Russo, and G.~Veneziano, ``{Universality of
		ultra-relativistic gravitational scattering},''
	\href{http://dx.doi.org/10.1016/j.physletb.2020.135924}{{\em Phys. Lett. B}
		{\bf 811} (2020)  135924}, \href{http://arxiv.org/abs/2008.12743}{{\tt
			arXiv:2008.12743 [hep-th]}}.
	
	\bibitem{Herrmann:2021lqe}
	E.~Herrmann, J.~Parra-Martinez, M.~S. Ruf, and M.~Zeng, ``{Gravitational
		Bremsstrahlung from Reverse Unitarity},''
	\href{http://dx.doi.org/10.1103/PhysRevLett.126.201602}{{\em Phys. Rev.
			Lett.} {\bf 126} (2021) no.~20, 201602},
	\href{http://arxiv.org/abs/2101.07255}{{\tt arXiv:2101.07255 [hep-th]}}.
	
	\bibitem{DiVecchia:2021bdo}
	P.~Di~Vecchia, C.~Heissenberg, R.~Russo, and G.~Veneziano, ``{The eikonal
		approach to gravitational scattering and radiation at $ \mathcal{O}
		$(G$^{3}$)},'' \href{http://dx.doi.org/10.1007/JHEP07(2021)169}{{\em JHEP}
		{\bf 07} (2021)  169}, \href{http://arxiv.org/abs/2104.03256}{{\tt
			arXiv:2104.03256 [hep-th]}}.
	
	\bibitem{Herrmann:2021tct}
	E.~Herrmann, J.~Parra-Martinez, M.~S. Ruf, and M.~Zeng, ``{Radiative classical
		gravitational observables at $ \mathcal{O} $(G$^{3}$) from scattering
		amplitudes},'' \href{http://dx.doi.org/10.1007/JHEP10(2021)148}{{\em JHEP}
		{\bf 10} (2021)  148}, \href{http://arxiv.org/abs/2104.03957}{{\tt
			arXiv:2104.03957 [hep-th]}}.
	
	\bibitem{Bjerrum-Bohr:2021vuf}
	N.~E.~J. Bjerrum-Bohr, P.~H. Damgaard, L.~Plant\'e, and P.~Vanhove,
	``{Classical gravity from loop amplitudes},''
	\href{http://dx.doi.org/10.1103/PhysRevD.104.026009}{{\em Phys. Rev. D} {\bf
			104} (2021) no.~2, 026009}, \href{http://arxiv.org/abs/2104.04510}{{\tt
			arXiv:2104.04510 [hep-th]}}.
	
	\bibitem{Bjerrum-Bohr:2021din}
	N.~E.~J. Bjerrum-Bohr, P.~H. Damgaard, L.~Plant\'e, and P.~Vanhove, ``{The
		amplitude for classical gravitational scattering at third Post-Minkowskian
		order},'' \href{http://dx.doi.org/10.1007/JHEP08(2021)172}{{\em JHEP} {\bf
			08} (2021)  172}, \href{http://arxiv.org/abs/2105.05218}{{\tt
			arXiv:2105.05218 [hep-th]}}.
	
	\bibitem{Brandhuber:2021eyq}
	A.~Brandhuber, G.~Chen, G.~Travaglini, and C.~Wen, ``{Classical gravitational
		scattering from a gauge-invariant double copy},''
	\href{http://dx.doi.org/10.1007/JHEP10(2021)118}{{\em JHEP} {\bf 10} (2021)
		118}, \href{http://arxiv.org/abs/2108.04216}{{\tt arXiv:2108.04216
			[hep-th]}}.
	
	\bibitem{Jakobsen:2022psy}
	G.~U. Jakobsen, G.~Mogull, J.~Plefka, and B.~Sauer, ``{All Things Retarded:
		Radiation-Reaction in Worldline Quantum Field Theory},''
	\href{http://arxiv.org/abs/2207.00569}{{\tt arXiv:2207.00569 [hep-th]}}.
	
	\bibitem{Kalin:2022hph}
	G.~K\"alin, J.~Neef, and R.~A. Porto, ``{Radiation-Reaction in the Effective
		Field Theory Approach to Post-Minkowskian Dynamics},''
	\href{http://arxiv.org/abs/2207.00580}{{\tt arXiv:2207.00580 [hep-th]}}.
	
	\bibitem{Weinberg:1964ew}
	S.~Weinberg, ``{Photons and Gravitons in $S$-Matrix Theory: Derivation of
		Charge Conservation and Equality of Gravitational and Inertial Mass},''
	\href{http://dx.doi.org/10.1103/PhysRev.135.B1049}{{\em Phys. Rev.} {\bf 135}
		(1964)  B1049--B1056}.
	
	\bibitem{Weinberg:1965nx}
	S.~Weinberg, ``{Infrared photons and gravitons},''
	\href{http://dx.doi.org/10.1103/PhysRev.140.B516}{{\em Phys. Rev.} {\bf 140}
		(1965)  B516--B524}.
	
	\bibitem{Weinberg:1972kfs}
	S.~Weinberg, {\em {Gravitation and Cosmology}: {Principles and Applications of
			the General Theory of Relativity}}.
	\newblock John Wiley and Sons, New York, 1972.
	
	\bibitem{Anastasiou:2002yz}
	C.~Anastasiou and K.~Melnikov, ``{Higgs boson production at hadron colliders in
		NNLO QCD},'' \href{http://dx.doi.org/10.1016/S0550-3213(02)00837-4}{{\em
			Nucl. Phys. B} {\bf 646} (2002)  220--256},
	\href{http://arxiv.org/abs/hep-ph/0207004}{{\tt arXiv:hep-ph/0207004}}.
	
	\bibitem{Anastasiou:2002qz}
	C.~Anastasiou, L.~J. Dixon, and K.~Melnikov, ``{NLO Higgs boson rapidity
		distributions at hadron colliders},''
	\href{http://dx.doi.org/10.1016/S0920-5632(03)80168-8}{{\em Nucl. Phys. B
			Proc. Suppl.} {\bf 116} (2003)  193--197},
	\href{http://arxiv.org/abs/hep-ph/0211141}{{\tt arXiv:hep-ph/0211141}}.
	
	\bibitem{Anastasiou:2003yy}
	C.~Anastasiou, L.~J. Dixon, K.~Melnikov, and F.~Petriello, ``{Dilepton rapidity
		distribution in the Drell-Yan process at NNLO in QCD},''
	\href{http://dx.doi.org/10.1103/PhysRevLett.91.182002}{{\em Phys. Rev. Lett.}
		{\bf 91} (2003)  182002}, \href{http://arxiv.org/abs/hep-ph/0306192}{{\tt
			arXiv:hep-ph/0306192}}.
	
	\bibitem{Veneziano:2022zwh}
	G.~Veneziano and G.~A. Vilkovisky, ``{Angular momentum loss in gravitational
		scattering, radiation reaction, and the Bondi gauge ambiguity},''
	\href{http://dx.doi.org/10.1016/j.physletb.2022.137419}{{\em Phys. Lett. B}
		{\bf 834} (2022)  137419}, \href{http://arxiv.org/abs/2201.11607}{{\tt
			arXiv:2201.11607 [gr-qc]}}.
	
	\bibitem{Bini:2012ji}
	D.~Bini and T.~Damour, ``{Gravitational radiation reaction along general orbits
		in the effective one-body formalism},''
	\href{http://dx.doi.org/10.1103/PhysRevD.86.124012}{{\em Phys. Rev. D} {\bf
			86} (2012)  124012}, \href{http://arxiv.org/abs/1210.2834}{{\tt
			arXiv:1210.2834 [gr-qc]}}.
	
	\bibitem{Bini:2021gat}
	D.~Bini, T.~Damour, and A.~Geralico, ``{Radiative contributions to
		gravitational scattering},''
	\href{http://dx.doi.org/10.1103/PhysRevD.104.084031}{{\em Phys. Rev. D} {\bf
			104} (2021) no.~8, 084031}, \href{http://arxiv.org/abs/2107.08896}{{\tt
			arXiv:2107.08896 [gr-qc]}}.
	
	\bibitem{Damour:2020tta}
	T.~Damour, ``{Radiative contribution to classical gravitational scattering at
		the third order in $G$},''
	\href{http://dx.doi.org/10.1103/PhysRevD.102.124008}{{\em Phys. Rev. D} {\bf
			102} (2020) no.~12, 124008}, \href{http://arxiv.org/abs/2010.01641}{{\tt
			arXiv:2010.01641 [gr-qc]}}.
	
	\bibitem{Kosower:2018adc}
	D.~A. Kosower, B.~Maybee, and D.~O'Connell, ``{Amplitudes, Observables, and
		Classical Scattering},''
	\href{http://dx.doi.org/10.1007/JHEP02(2019)137}{{\em JHEP} {\bf 02} (2019)
		137},
	\href{http://arxiv.org/abs/1811.10950}{{\tt arXiv:1811.10950 [hep-th]}}.
	
	\bibitem{KoemansCollado:2019ggb}
	A.~Koemans~Collado, P.~Di~Vecchia, and R.~Russo, ``{Revisiting the second
		post-Minkowskian eikonal and the dynamics of binary black holes},''
	\href{http://dx.doi.org/10.1103/PhysRevD.100.066028}{{\em Phys. Rev. D} {\bf
			100} (2019) no.~6, 066028}, \href{http://arxiv.org/abs/1904.02667}{{\tt
			arXiv:1904.02667 [hep-th]}}.
	
	\bibitem{Cristofoli:2020uzm}
	A.~Cristofoli, P.~H. Damgaard, P.~Di~Vecchia, and C.~Heissenberg,
	``{Second-order Post-Minkowskian scattering in arbitrary dimensions},''
	\href{http://dx.doi.org/10.1007/JHEP07(2020)122}{{\em JHEP} {\bf 07} (2020)
		122}, \href{http://arxiv.org/abs/2003.10274}{{\tt arXiv:2003.10274
			[hep-th]}}.
	
	\bibitem{Goldberger:2016iau}
	W.~D. Goldberger and A.~K. Ridgway, ``{Radiation and the classical double copy
		for color charges},''
	\href{http://dx.doi.org/10.1103/PhysRevD.95.125010}{{\em Phys. Rev. D} {\bf
			95} (2017) no.~12, 125010}, \href{http://arxiv.org/abs/1611.03493}{{\tt
			arXiv:1611.03493 [hep-th]}}.
	
	\bibitem{Luna:2017dtq}
	A.~Luna, I.~Nicholson, D.~O'Connell, and C.~D. White, ``{Inelastic Black Hole
		Scattering from Charged Scalar Amplitudes},''
	\href{http://dx.doi.org/10.1007/JHEP03(2018)044}{{\em JHEP} {\bf 03} (2018)
		044}, \href{http://arxiv.org/abs/1711.03901}{{\tt arXiv:1711.03901
			[hep-th]}}.
	
	\bibitem{Mogull:2020sak}
	G.~Mogull, J.~Plefka, and J.~Steinhoff, ``{Classical black hole scattering from
		a worldline quantum field theory},''
	\href{http://dx.doi.org/10.1007/JHEP02(2021)048}{{\em JHEP} {\bf 02} (2021)
		048}, \href{http://arxiv.org/abs/2010.02865}{{\tt arXiv:2010.02865
			[hep-th]}}.
	
	\bibitem{Kulish:1970ut}
	P.~P. Kulish and L.~D. Faddeev, ``{Asymptotic conditions and infrared
		divergences in quantum electrodynamics},''
	\href{http://dx.doi.org/10.1007/BF01066485}{{\em Theor. Math. Phys.} {\bf 4}
		(1970)  745}.
	
	\bibitem{Choi:2017ylo}
	S.~Choi and R.~Akhoury, ``{BMS Supertranslation Symmetry Implies Faddeev-Kulish
		Amplitudes},'' \href{http://dx.doi.org/10.1007/JHEP02(2018)171}{{\em JHEP}
		{\bf 02} (2018)  171}, \href{http://arxiv.org/abs/1712.04551}{{\tt
			arXiv:1712.04551 [hep-th]}}.
	
	\bibitem{Mirbabayi:2016axw}
	M.~Mirbabayi and M.~Porrati, ``{Dressed Hard States and Black Hole Soft
		Hair},'' \href{http://dx.doi.org/10.1103/PhysRevLett.117.211301}{{\em Phys.
			Rev. Lett.} {\bf 117} (2016) no.~21, 211301},
	\href{http://arxiv.org/abs/1607.03120}{{\tt arXiv:1607.03120 [hep-th]}}.
	
	\bibitem{Hannesdottir:2019opa}
	H.~Hannesdottir and M.~D. Schwartz, ``{$S$ -Matrix for massless particles},''
	\href{http://dx.doi.org/10.1103/PhysRevD.101.105001}{{\em Phys. Rev. D} {\bf
			101} (2020) no.~10, 105001}, \href{http://arxiv.org/abs/1911.06821}{{\tt
			arXiv:1911.06821 [hep-th]}}.
	
	\bibitem{Bonga:2018gzr}
	B.~Bonga and E.~Poisson, ``{Coulombic contribution to angular momentum flux in
		general relativity},''
	\href{http://dx.doi.org/10.1103/PhysRevD.99.064024}{{\em Phys. Rev. D} {\bf
			99} (2019) no.~6, 064024}, \href{http://arxiv.org/abs/1808.01288}{{\tt
			arXiv:1808.01288 [gr-qc]}}.
	
	\bibitem{DeTar:1971pmj}
	C.~E. DeTar, D.~Z. Freedman, and G.~Veneziano, ``{Sum rules for inclusive
		cross-sections},'' \href{http://dx.doi.org/10.1103/PhysRevD.4.906}{{\em Phys.
			Rev. D} {\bf 4} (1971)  906--909}.
	
	\bibitem{DiVecchia:2021ndb}
	P.~Di~Vecchia, C.~Heissenberg, R.~Russo, and G.~Veneziano, ``{Radiation
		Reaction from Soft Theorems},''
	\href{http://dx.doi.org/10.1016/j.physletb.2021.136379}{{\em Phys. Lett. B}
		{\bf 818} (2021)  136379}, \href{http://arxiv.org/abs/2101.05772}{{\tt
			arXiv:2101.05772 [hep-th]}}.
	
	\bibitem{Gralla:2021eoi}
	S.~E. Gralla and K.~Lobo, ``{Electromagnetic scoot},''
	\href{http://dx.doi.org/10.1103/PhysRevD.105.084053}{{\em Phys. Rev. D} {\bf
			105} (2022) no.~8, 084053}, \href{http://arxiv.org/abs/2112.01729}{{\tt
			arXiv:2112.01729 [gr-qc]}}.
	
	\bibitem{Gralla:2021qaf}
	S.~E. Gralla and K.~Lobo, ``{Self-force effects in post-Minkowskian
		scattering},'' \href{http://dx.doi.org/10.1088/1361-6382/ac5d88}{{\em Class.
			Quant. Grav.} {\bf 39} (2022) no.~9, 095001},
	\href{http://arxiv.org/abs/2110.08681}{{\tt arXiv:2110.08681 [gr-qc]}}.
	
	\bibitem{Dlapa:2022lmu}
	C.~Dlapa, G.~K\"alin, Z.~Liu, J.~Neef, and R.~A. Porto, ``{Radiation Reaction
		and Gravitational Waves at Fourth Post-Minkowskian Order},''
	\href{http://arxiv.org/abs/2210.05541}{{\tt arXiv:2210.05541 [hep-th]}}.
	
	\bibitem{Bini:2022enm}
	D.~Bini, T.~Damour, and A.~Geralico, ``{Radiated momentum in gravitational
		two-body scattering including time-asymmetric effects},''
	\href{http://arxiv.org/abs/2210.07165}{{\tt arXiv:2210.07165 [gr-qc]}}.
	
	\bibitem{toap2}
	E.~Herrmann, A.~Moss, J.~Parra-Martinez, and M.~Ruf (in preparation).
	
	\bibitem{Britto:2021pud}
	R.~Britto, R.~Gonzo, and G.~R. Jehu, ``{Graviton particle statistics and
		coherent states from classical scattering amplitudes},''
	\href{http://dx.doi.org/10.1007/JHEP03(2022)214}{{\em JHEP} {\bf 03} (2022)
		214}, \href{http://arxiv.org/abs/2112.07036}{{\tt arXiv:2112.07036
			[hep-th]}}.
	
\end{thebibliography}

\providecommand{\href}[2]{#2}\begingroup\raggedright\endgroup

\end{document}